\documentclass[aip,jmp,msc,preprint]{revtex4-1}
\usepackage{eurosym}
\usepackage{graphicx}
\usepackage{amsmath}
\usepackage{bm}
\usepackage{color}
\usepackage{dcolumn}
\usepackage{bm}

\def\be{\begin{equation}}
\def\ee{\end{equation}}
\def\bea{\begin{eqnarray}}
\def\eea{\end{eqnarray}}

\begin{document}

\title{Exact travelling wave solutions of non-linear
reaction-convection-diffusion equations -- an Abel equation based approach}
\author{T. Harko}\affiliation{Department of Mathematics, University College London, Gower Street, London
WC1E 6BT, United Kingdom}
\email{t.harko@ucl.ac.uk}
\author{M. K. Mak}\affiliation{Departamento de  F\'{i}sica, Facultad de Ciencias Naturales, Universidad de Atacama, Copayapu 485, Copiap\'{o}, Chile.
}
\email{mkmak@uda.cl}

\begin{abstract}
We consider quasi-stationary (travelling wave type) solutions to a general
nonlinear reaction-convection-diffusion equation with arbitrary, autonomous
coefficients. The second order nonlinear equation describing one dimensional
travelling waves can be reduced to a first kind first order Abel type differential equation.
By using two integrability conditions for the Abel equation (the Chiellini
lemma and the Lemke transformation), several classes of exact travelling
wave solutions of the general reaction--convection-diffusion equation are
obtained, corresponding to different functional relations imposed between
the diffusion, convection and reaction functions. In particular, we obtain
travelling wave solutions for two non-linear second order partial
differential equations, representing generalizations of the standard diffusion equation and of the classical
Fisher--Kolmogorov equation, to which they reduce for some limiting values of the model parameters. The models correspond to some specific, power law type choices of the reaction and convection functions, respectively. The travelling wave solutions of these two classes of differential equations are investigated in detail by using both numerical and semi-analytical methods.
\end{abstract}

\maketitle
%\pacs{67.85.Jk, 04.40.Dg, 95.30.Cq, 95.30.Sf}

\section{Introduction}

Reaction--convection--diffusion equations, representing, from a mathematical
point of view, a particular class of second order non-linear partial
differential equations, play a very important role in many areas of physics,
chemistry, engineering and biology. Different phenomena, such as heat
transfer, combustion, reaction chemistry, fluid dynamics, plasma physics,
soil-moisture, foam drainage, crystal growth, biological population
genetics, cellular ecology, neurology, synergy etc. can be described
mathematically by using reaction - diffusion equations \cite{9cm}-\cite{b5}.
A classic example of a reaction - diffusion equations is the
Fisher--Kolmogorov--Petrovskii--Piscunov equation (Fisher--Kolmogorov for
short) \cite{Fish}-\cite{Kolm},
\begin{equation}\label{fk}
\frac{\partial u(t,x)}{\partial t}=\frac{\partial ^{2}u(t,x)}{\partial x^{2}}%
+u(t,x)\left[ 1-u(t,x)\right] ,
\end{equation}%
which is the archetypical deterministic model for the spread of an
advantageous gene in a population of diploid individuals living in a
one-dimensional habitat \cite{13}. The Newell--Whitehead equation or
amplitude equation \cite{N_W,S},
\begin{equation*}
\frac{\partial u(t,x)}{\partial t}=\frac{\partial ^{2}u(t,x)}{\partial x^{2}}%
+u(t,x)\left[ 1-u^{2}(t,x)\right] ,
\end{equation*}%
arises after carrying out a suitable normalization in the study of thermal
convection of a fluid heated from below. Considering the perturbation from a
stationary state, the equation describes the evolution of the amplitude of
the vertical velocity, if this varies slowly. The Zeldovich equation \cite%
{Zel,Zel1},
\begin{equation*}
\frac{\partial u(t,x)}{\partial t}=\frac{\partial ^{2}u(t,x)}{\partial x^{2}}%
+u^{2}(t,x)\left[ 1-u(t,x)\right] ,
\end{equation*}%
has many applications in combustion theory. The unknown $u$ represents
temperature, while the last term on the right-hand side corresponds to the
generation of heat by combustion. The FitzHugh--Nagumo equation or the
bistable equation \cite{Fitz,Nag},
\begin{equation*}
\frac{\partial u(t,x)}{\partial t}=\frac{\partial ^{2}u(t,x)}{\partial x^{2}}%
+u(t,x)\left[ u(t,x)-\alpha \right] ,0<\alpha <1,
\end{equation*}%
models the transmission of electrical pulses in a nerve axon.

All of the above equations, and many more besides, are specific examples of
equations of the general class of reaction-convection-diffusion equations
\cite{9cm,b3},
\begin{equation}
\frac{\partial u(t,x)}{\partial t}=\frac{\partial }{\partial x}\left[ D(u)%
\frac{\partial u(t,x)}{\partial x}\right] +B(u)\frac{\partial u(t,x)}{%
\partial x}+Q(u),  \label{1}
\end{equation}%
where the function $D(u)$ plays the role of the diffusion coefficient, $B(u)$
may be viewed as a nonlinear convective flux function, and $Q(u)$ is the
reaction function, respectively.

A class of non-linear second order partial differential equations, which
embodies a number of well known reaction--convection--diffusion equations,
is given by \cite{9cm,b3}
\begin{equation}\label{gen}
\frac{\partial u}{\partial t}=\frac{\partial ^{2}u^{m}}{\partial x^{2}}+%
\frac{\partial }{\partial x}\left[ \left( b_{0}+b_{1}u^{p}\right) u\right]
+u^{2-m}\left( 1-u^{p}\right) \left( c_{0}+c_{1}u^{p}\right) ,u>0,
\end{equation}%
where $m$, $p$, $b_{0}$, $b_{1}$, $c_{0}$ and $c_{1}$ are constants. The
Fisher, or the logistic equation, the Newell-Whitehead, or amplitude
equation, the Zeldovich equation, the Nagumo or the bistable equation, are
all particular cases of Eq.~(\ref{gen}), corresponding to different values of the constant parameters.

Reaction-diffusion equations usually do admit travelling wave solutions \cite%
{9cm}-\cite{b5}. There are many observations and experiments showing the
presence of wave phenomena in a large number of natural reaction, convection
and diffusion processes. This alone represents a very powerful motivation
for studying their occurrence. Other reasons why the study of
travelling-wave solutions has become such an important part of the
mathematical analysis of nonlinear reaction--convection--diffusion processes
have been summarized in \cite{b3}, and are that: the analysis of travelling
waves provides an easy mean of finding explicit solutions of the equation;
in general travelling-wave solutions are easier to analyze and therewith
discern properties to be expected of other solutions; travelling wave
solutions can be used as tools in comparison principles and the like to
determine the properties of general solutions; and, last but not least, in
conformance with their natural occurrence in many mathematically modelled
phenomena, in numerous situations they characterize the long-term behavior
of the solutions \cite{b3}.

There is a very large literature on the travelling wave solutions in general
non-linear reaction--convection--diffusion equations, and on their
scientific applications. Wave solutions, which correspond to moving bands of
the concentration of the chemical constituents do exist in the system of
equations describing the Belousov-Zhabotinskii reaction \cite{9c1}. By using
parameter values obtained from experimental studies, a numerical analysis
has been performed to study the chemical travelling wave solutions, and
different relevant properties of the solutions have also been obtained. A
semi-inverse method, which allows to obtain exact static solutions of
one-component, one-dimensional reaction-diffusion equations with variable
diffusion function $D$, and which requires at most qualitative information
on its spatial dependence, was introduced in \cite{9c01}. By using a simple
ansatz, the reaction-diffusion equations can be mapped onto the stationary
Schr\"{o}dinger equations, with the general form of the potential
undetermined. The problem of finding quasi-stationary solutions to a
nonlinear reaction-diffusion equation with arbitrary, autonomous
coefficients was considered, by using a new substitution, in \cite{9c2}. By
solving the corresponding Abel equations exact periodical and solitary wave
solutions were obtained. As shown in \cite{9c2}, the initial general
reaction--diffusion equation can be reduced to either a linear ordinary
differential equation, or to the first order first kind Abel differential
equation, of the form
\begin{equation*}
y^{\prime }=f_0(x)+f_1(x)y+f_2(x)y^2+f_3(x)y^3,
\end{equation*}
where $f_i(x)$, $i=0,1,2,3$, are arbitrary real functions of $x$, defined on
a real interval $I\subseteq \Re $, with $f_0,f_1,f_2,f_3\in C^{\infty }(I)$
\cite{Golub}.

A direct method to obtain travelling-wave solutions of some nonlinear
partial differential equations, expressed in terms of solutions of the first
order Abel differential equation of the first kind, with constant
coefficients, was proposed in \cite{9c3}. Travelling wave solutions have
also many important technological applications. In particular, it was shown
that the theoretical problem of the kinetics of thin film (or of wire-like
nanostructures) growth has bounded solutions expressed in terms of elliptic
functions \cite{9c3}. Exact solutions to the viscosity modified Benjamin,
Bona, and Mahony (BBM) equation were found and studied in \cite{9c4}, by
including the effect of a small dissipation on the travelling waves. Using
Lyapunov functions and dynamical systems theory, it was proven that when
viscosity is added to the BBM equation, in certain regions there still exist
bounded travelling wave solutions in the form of solitary waves, periodic,
and elliptic functions.

Travelling-wave solutions of different mathematical model describing the
growth of tumors have been considered in several publications \cite{9c}-\cite%
{H14}. Spreading cell fronts play an essential role in many physiological
and biological processes. Classically, models of this process are based on
the Fisher - Kolmogorov equation \cite{Fish, Fish1, Kolm,13}; however, in
many biological situations such continuum representations are not always
suitable, as they do not explicitly represent behavior at the level of
individual cells. Additionally, many models examine only the large time
asymptotic behavior, where a travelling wave front with a constant speed has
been established \cite{9d}.

It is the goal of the present paper to study exact traveling wave solutions
for the one-dimensional general reaction--convection-diffusion Eq.~(\ref{1}%
). By considering travelling wave solutions of Eq.~(\ref{1}), of the form $%
u=u(x-V_{f}t)=u(\xi )$, where $V_{f}=\mathrm{constant}$, we show, as a first
step in our study, that the second order non-linear differential equation
describing wave propagation can be reduced to a first kind first order
non-linear Abel equation \cite{kamke}. The mathematical properties of the
Abel equation and its applications in different fields have been considered
recently in \cite{18}-\cite{Rosu4}.

By using some standard integrability conditions of the Abel equation, we
obtain several classes of exact travelling wave solutions of the general
reaction--convection--diffusion Eq.~(\ref{1}). More exactly, the first class
of solutions is obtained by using the integrability condition of Chiellini
\cite{kamke, Chiel}, which can be formulated as a differential condition
relating the diffusion, convection and reaction functions. The use of this
condition allows us to obtain exact travelling wave solutions to some
reaction--convection--diffusion equations, representing a generalization of
the diffusion and Fisher--Kolmogorov equations, for some specific forms of
the functions $D(u)$, $Q(u)$ and $B(u)$. More exactly, we consider the cases
for which
\begin{equation*}
D(u)\propto \left( 1-u/u_{max}\right) ^{-\alpha },Q(u)\propto \left(
1-u/u_{max}\right) ^{\alpha },B(u)=\mathrm{constant},
\end{equation*}%
where $\alpha $ and $u_{max}$ are constants, and we present the closed form
solution of the corresponding reaction--convection--diffusion equation. The
case $\alpha =1$ is investigated in detail. For these choices of $D(u)$ and $%
Q(u)$ the most general form of the convection function $B(u)$ that allows
exact travelling wave solutions is also obtained.

The second class of solutions is obtained by transforming the Abel equation
to a second order non-linear differential equation, by using a method
introduced in \cite{Lemke} (the Lemke transformation). This transformation
also allows the construction of exact travelling wave solutions for general
reaction--convection--diffusion equations for several distinct choices of
the diffusion, convection and reaction functions.

As specific numerical examples of the travelling wave propagation in second
order non-linear reaction--diffusion--convection partial differential
equations we consider two particular cases, in which the diffusion and
reaction functions are represented in a power law form. These two equations
represent generalizations of the standard diffusion and Fisher--Kolmogorov
equations. After being transformed to a dimensionless form they are
investigated in detail numerically, and the corresponding evolution profile
of the travelling waves is obtained.

The present paper is organized as follows. In Section~\ref{sect2} we present
the reduction of the ordinary second order travelling wave differential
equation, associated to the general reaction--convection--diffusion second
order partial differential equation to a first kind first order Abel
ordinary differential equation. For specific choices of the reaction,
diffusion and convection functions the solution of the Abel equation can be
obtained in an exact form with the use of the Chiellini integrability
condition. Several classes of exact travelling wave solutions are obtained.
A second integrability case for the Abel equation, based on the Lemke
transformation, is used in Section~\ref{sect3} to construct several exact
travelling wave solutions of the general reaction--convection--diffusion
equation, with the coefficients satisfying particular functional relations.
In Section~\ref{sect4} we numerically investigate the properties of the
travelling wave solutions, corresponding to two generalizations of the
Fisher--Kolmogorov equation. We discuss and conclude our results in Section~%
\ref{sect5}.

\section{Exact travelling wave solutions of the general
reaction-convection-diffusion equation - the Chiellini lemma based approach}

\label{sect2}

In the following we obtain exact travelling wave solutions of the 1+1
dimensional general reaction - convection - diffusion Eq.~(\ref{1}) where
the dissipative (diffusion) function $D(u)\neq 0$, the convection function $%
B(u)\neq 0$ and the adsorption (or reaction) term $Q(u)\neq 0$ all depend
explicitly upon the function $u$ only, and not on space $x$ and time $t$
variables. As a first step in our study we reduce the basic equation
describing travelling wave solutions to a first order first kind Abel
equation. By using the Chiellini integrability condition we obtain several
classes of exact travelling wave solutions of the general
reaction--convection--diffusion equation, corresponding to different
functional relations between $D(u)$, $Q(u)$ and $B(u)$.

\subsection{Reduction to the Abel equation}

By introducing a phase variable $\xi =x-V_{f}t$, where $V_{f}=\mathrm{%
constant}\geq 0$ is a constant wave velocity, Eq.~(\ref{1}) takes the form
of a second order non-linear differential equation of the form
\begin{equation}
\frac{d^{2}u}{d\xi ^{2}}+\alpha (u)\left( \frac{du}{d\xi }\right) ^{2}+\beta
(u)\frac{du}{d\xi }+\gamma (u)=0,  \label{3}
\end{equation}%
where
\begin{equation*}
\alpha (u)=\frac{d}{du}\ln D(u),\beta (u)=\frac{V_{f}+B(u)}{D(u)},\gamma (u)=%
\frac{Q(u)}{D(u)}.
\end{equation*}%
The second order non-linear differential Eq.~(\ref{3}) must be considered
together with the initial conditions $u(0)=u_{0}$ and $\left. \left( du/d\xi
\right) \right\vert _{\xi =0}=u_{0}^{\prime }$, respectively.

By means of the transformations
\begin{equation}
\frac{du}{d\xi }=\sigma ,\frac{d^{2}u}{d\xi ^{2}}=\sigma \frac{d\sigma }{du}%
,\sigma =\frac{1}{v},  \label{k1}
\end{equation}%
Eq.~(\ref{3}) can be transformed to the general form of the first order
first kind Abel equation, given by
\begin{equation}
\frac{dv}{du}=\alpha (u)v+\beta (u)v^{2}+\gamma (u)v^{3}.  \label{4}
\end{equation}%
Eq.~(\ref{4}) must be integrated with the initial condition $v\left(
u_{0}\right) =1/\left. \left( du/d\xi \right) \right\vert
_{u=u_{0}}=1/u_{0}^{\prime }$.

By introducing a new variable $w$, defined as
\begin{equation*}
v=e^{\int {\alpha (u)du}}w=D(u)w,  \label{kkk}
\end{equation*}%
allows us to write Eq.~(\ref{4}) in the standard form of the first kind Abel
equation,
\begin{equation}
\frac{dw}{du}=f(u)w^{2}+g(u)w^{3},  \label{5}
\end{equation}%
where we have denoted
\begin{equation*}
f(u)=V_{f}+B(u),
\end{equation*}%
and
\begin{equation*}
g(u)=D(u)Q(u),
\end{equation*}%
respectively. Eq.~(\ref{5}) must be solved with the initial condition
\begin{equation*}
w\left( u_{0}\right) =w_{0}=\frac{1}{D\left( u_{0}\right) u_{0}^{\prime }}.
\label{inw}
\end{equation*}

\subsection{The Chiellini integrability condition for general
reaction--convection--diffusion systems}

An exact integrability condition for the Abel equation Eq.~(\ref{5}) was
obtained by Chiellini \cite{Chiel} (see also \cite{kamke} and \cite{HLM}),
and can be formulated as the Chiellini Lemma as follows: %\begin{lemma}

\textbf{Chiellini Lemma}. \textit{If the coefficients $f(u)$ and $g(u)$ of a
first kind Abel type differential equation of the form given by Eq.~(\ref{5})%
} %\begin{equation}\label{abels}
%\frac{dw}{du}=f(u)w^{2}+g(u)w^{3}.
%\end{equation}%
\textit{\ satisfy the condition }%
\begin{equation}  \label{15}
\frac{d}{du}\left[ \frac{g(u)}{f(u)}\right] =kf(u),
\end{equation}
\textit{where $k=\mathrm{constant}\neq 0$, then the first kind Abel Eq.~(\ref%
{5}) can be exactly integrated}. %\end{lemma}

In order to prove the Chiellini Lemma, we introduce a new dependent variable
$\theta $, defined as \cite{Chiel, kamke}
\begin{equation}
w=\frac{f(u)}{g(u)}\theta .  \label{kk}
\end{equation}

Then Eq.~(\ref{5}) can be written as
\begin{equation}
\left[ \frac{1}{g(u)}\frac{df(u)}{du}-\frac{f(u)}{g^{2}(u)}\frac{dg(u)}{du}%
\right] \theta +\frac{f(u)}{g(u)}\frac{d\theta }{du}=\frac{f^{3}(u)}{g^{2}(u)%
}\left( \theta ^{3}+\theta ^{2}\right) .  \label{6}
\end{equation}%
On the other hand, the condition given by Eq.~(\ref{15}) can be written in
an equivalent form as
\begin{equation}
\frac{f(u)}{g^{2}(u)}\frac{dg(u)}{du}-\frac{1}{g(u)}\frac{df(u)}{du}=k\frac{%
f^{3}(u)}{g^{2}(u)}.
\end{equation}%
Therefore Eq.~(\ref{6}) becomes
\begin{equation}
\frac{d\theta }{du}=\frac{f^{2}(u)}{g(u)}\theta \left( \theta ^{2}+\theta
+k\right) ,  \label{6_1}
\end{equation}%
which is a first order separable differential equation, with the general
solution given by
\begin{equation}  \label{intc}
\int {\frac{f^{2}(u)}{g(u)}du}=\int {\frac{d\theta }{\theta \left( \theta
^{2}+\theta +k\right) }}\equiv \frac{F(\theta,k)}{k}.
\end{equation}

With the use of the condition (\ref{15}), the left hand side of Eq.~(\ref%
{intc}) can be written as \cite{Rosu1}
\begin{equation}
\int {\frac{f^{2}(u)}{g(u)}du}=\frac{1}{k}\int {\frac{d}{du}\ln }\left\vert
\frac{g(u)}{f(u)}\right\vert {du}=\frac{1}{k}\ln \left\vert \frac{g(u)}{f(u}%
\right\vert +C_{0},
\end{equation}%
where $C_{0}$ is an arbitrary constant of integration. Therefore the general
solution of Eq.~(\ref{6_1}) is obtained as
\begin{equation}
\frac{g(u)}{f(u)}=C^{-1}e^{F(\theta ,k)},  \label{sol}
\end{equation}%
where $C^{-1}=\exp \left( -kC_{0}\right) $ is an arbitrary constant of
integration, and
\begin{equation}  \label{sol2}
e^{F(\theta ,k)}=\left\{
\begin{array}{lll}
\frac{\theta }{\sqrt{\theta ^{2}+\theta +k}}\exp \left[ -\frac{1}{\sqrt{4k-1}%
}\arctan \left( \frac{1+2\theta }{\sqrt{4k-1}}\right) \right] ,\qquad k>%
\frac{1}{4}, &  &  \\
&  &  \\
\frac{\theta }{1+2\theta }e^{1/(1+2\theta )}, \qquad k=\frac{1}{4}, &  &  \\
&  &  \\
\frac{\theta }{\sqrt{\theta ^{2}+\theta +k}}\exp \left[ \frac{1}{\sqrt{1-4k}}%
\mathrm{arctanh}\left( \frac{1+2\theta }{\sqrt{1-4k}}\right) \right] ,\qquad
k<1/4, &  &
\end{array}%
\right.
\end{equation}
respectively. Eq.~(\ref{sol}) determines $\theta $ as a function of $u$.

%The Chiellini integrability condition given by Eq.~(\ref{15}) can be written
%in an equivalent form as
%\begin{equation}
%\frac{dg(u)}{du}=\frac{1}{f(u)}\frac{df(u)}{du}g(u)+kf^2(u),
%\end{equation}
%representing a first order linear differential equation in $g(u)$. As a
%function of $g(u)$, the function $f(u)$ satisfies the following first order
%differential equation
%\begin{equation}  \label{c22}
%\frac{1}{f(u)}\frac{df(u)}{du}=-k\frac{1}{g(u)}f^2(u)+\frac{1}{g(u)}\frac{%
%dg(u)}{du}.
%\end{equation}

%In order to solve Eq.~(\ref{c22}) we introduce a new dependent variable $%
%f(u)=1/\sigma (u)$, and by denoting $\sigma ^2(x)=\zeta (x)$, we obtain a
%first order differential equation for $\zeta $,
%\begin{equation}
%\frac{d\zeta (u)}{du}=-2\frac{1}{g(u)}\frac{dg(u)}{du}\zeta (u)+\frac{2k}{%
%g(u)}.
%\end{equation}

The \textit{Chiellini integrability Theorem} for the general
reaction--convection--diffusion Eq.~(\ref{1}) is given as follows.

\textbf{Theorem 1.} \textit{If the reaction, convection and diffusion
functions $Q(u)$, $B(u)$ and $D(u)$ of the general
reaction-convection-diffusion Eq.~(\ref{1}) satisfy the conditions
\begin{equation}
D(u)Q(u)=\left[ V_{f}+B(u)\right] \left[ C_{1}+kV_{f}u+k\int {B(u)du}\right]
,  \label{th11}
\end{equation}%
or
\begin{equation}
V_{f}+B(u)=\pm \frac{D(u)Q(u)}{\sqrt{C_{2}+2k\int {D(u)Q(u)du}}},
\label{th12}
\end{equation}%
where $C_{1}$, $C_{2}$, and $k$ are arbitrary constants, then the reaction -
convection - diffusion Eq.~(\ref{1}) admits exact travelling wave solutions $%
u(\xi )$, with $\xi =x-V_{f}t$, given in a parametric form, with $\theta $
taken as a parameter, by
\begin{equation}
\xi (\theta )-\xi _{0}\left( \theta _{0}\right) =\int_{\theta _{0}}^{\theta }%
{\frac{D\left[ u(\psi )\right] d\psi }{\left\{ V_{f}+B[u(\psi )]\right\}
\left( \psi ^{2}+\psi +k\right) }},u=u(\theta ),  \label{k4}
\end{equation}%
where $u(\theta )$ is obtained as a solution of the equation
\begin{equation}
\frac{D(u)Q(u)}{V_{f}+B(u)}=C^{-1}e^{F(\theta ,k)},  \label{k5}
\end{equation}%
and with the functions $F(\theta ,k)$ given by Eqs.~(\ref{sol2}). }

\textbf{Proof of Theorem 1}. As applied to the Abel Eq.~(\ref{5}),
describing the travelling wave solutions of the general
reaction--convection--diffusion Eq.~(\ref{1}), the Chiellini lemma states that
if the coefficients of the equation satisfy the condition
\begin{equation}
\frac{d}{du}\left[ \frac{D(u)Q(u)}{V_{f}+B(u)}\right] =k\left[ V_{f}+B(u)%
\right] ,  \label{cond}
\end{equation}%
Eq.~(\ref{5}) is exactly integrable. By integrating Eq.~(\ref{cond}) we
obtain immediately Eq.~(\ref{th11}), representing a first integrability
condition for obtaining travelling wave solutions of Eq.~(\ref{1}).
Alternatively, by introducing the notation $Y(u)=V_{f}+B(u)$, we can write
the integrability condition Eq.~(\ref{cond}) in the equivalent form
\begin{equation*}
\frac{dY(u)}{du}=\frac{d}{du}\ln \left[ D(u)Q(u)\right] Y(u)-\frac{k}{%
D(u)Q(u)}Y^{3}(u).
\end{equation*}%
By dividing the above equation by $Y^{3}(u)$, and denoting $Z(u)Y^{2}(u)=1$,
we obtain for $Z(u)$ the first order linear equation
\begin{equation*}
\frac{dZ(u)}{du}=-2\frac{d}{du}\ln \left[ D(u)Q(u)\right] Z(u)+\frac{2k}{%
D(u)Q(u)},
\end{equation*}%
with the general solution given by
\begin{equation*}
Z(u)=\frac{1}{\left[ V_{f}+B(u)\right] ^{2}}=\frac{1}{\left[ D(u)Q(u)\right]
^{2}}\left[ C_{2}+2k\int {D(u)Q(u)du}\right] ,
\end{equation*}%
where $C_{2}$ is an arbitrary constant of integration, which gives
immediately the second integrability condition (\ref{th12}) for the general
reaction--convection--diffusion Eq.~(\ref{1}).

From Eqs.~(\ref{k1}) and (\ref{kk}) we obtain the relation
\begin{equation*}
\frac{du}{d\xi }=\frac{1}{v}=\frac{1}{D(u)w(u)}=\frac{Q(u)}{\left[V_f+B(u)%
\right]\theta }.
\end{equation*}
From Eq.~(\ref{6_1}) we obtain
\begin{equation*}
\frac{du}{d\theta }=\frac{du}{d\xi}\frac{d\xi}{d\theta }=\frac{D(u)Q(u)}{%
\left[V_f+B(u)\right]^2\theta \left(\theta ^2+\theta +k\right)},
\end{equation*}
giving
\begin{equation*}
\frac{d\xi }{d\theta}=\frac{D(u)}{\left[V_f+B(u)\right]\left(\theta
^2+\theta +k\right)}.
\end{equation*}

By integrating the above equation we obtain the parametric dependence of $%
\xi $ on $\theta $ as given by Eq.~(\ref{k4}). From Eq.~(\ref{sol}) we
obtain the parametric dependence of $u$ as a function of the parameter $%
\theta $, as given in Eq.~(\ref{k5}). This concludes the proof of \textbf{%
Theorem 1}.

\subsection{Travelling wave solutions for $D(u)=\mathrm{constant}$, $B(u)=%
\mathrm{constant}$ and $Q(u)\propto u$}

As a first example of the application of \textbf{Theorem 1} we consider the
simple case when the diffusion convection and the reaction functions are
given by $D(u)=D_{0}=\mathrm{constant}$, $B(u)=B_{0}=\mathrm{constant}$, and
$Q(u)=\rho u$, $\rho =\mathrm{constant}$, respectively. Therefore the
equation describing the travelling wave propagation in the corresponding
reaction-convection-diffusion system takes the form
\begin{equation}
\frac{d^{2}u}{d\xi ^{2}}+\frac{V_{f}+B_{0}}{D_{0}}\frac{du}{d\xi }+\frac{%
\rho }{D_{0}}u=0.  \label{ex1}
\end{equation}%
The integrability condition given by Eq.~(\ref{cond}) fixes the constant $k$
as
\begin{equation*}
k=\frac{\rho D_{0}}{\left( V_{f}+B_{0}\right) ^{2}}.
\end{equation*}%
Then from Eqs.~(\ref{k5}) of \textbf{Theorem 1} we obtain the function $%
u(\theta )$ as %\begin{equation}
%\xi (\theta )-\xi _{0}=-\frac{2D_{0}}{\Delta }\tanh ^{-1}\left[ \frac{2D_{0}\rho
%\theta+\left(V_{f}+B_0\right)^{2}}{\left(V_{f}+B_0\right)\Delta }\right] ,\frac{\rho D_{0}}{\left(V_{f}+B_0\right)^{2}}\neq \frac{1}{4},
%\end{equation}%
\begin{equation*}  \label{33}
u(\theta )=\frac{V_{f}+B_{0}}{D_{0}\rho }C^{-1}e^{F(\theta ,k)},\frac{\rho
D_{0}}{\left( V_{f}+B_{0}\right) ^{2}}\neq \frac{1}{4},
\end{equation*}%
\begin{equation}  \label{34}
u(\theta )=\frac{V_{f}+B_{0}}{D_{0}\rho }C^{-1}\frac{\theta }{2\theta +1}e^{%
\frac{1}{2\theta +1}},\frac{\rho D_{0}}{\left( V_{f}+B_{0}\right) ^{2}}=%
\frac{1}{4}.
\end{equation}

The parametric dependence of $\xi $ is obtained with the help of Eq.~(\ref%
{k4}), which gives
\begin{equation*}
\left( \xi -\xi _{0}\right) \frac{V_{f}+B_{0}}{D_{0}}=\int \frac{d\theta }{%
\theta ^{2}+\theta +k}=\int \frac{d\theta }{\left( \theta +\frac{1}{2}%
\right) ^{2}-\left[ \frac{\Delta }{2\left( V_{f}+B_{0}\right) }\right] ^{2}},%
\frac{\rho D_{0}}{\left( V_{f}+B_{0}\right) ^{2}}\neq \frac{1}{4},
\end{equation*}
where we have denoted
\begin{equation*}
\Delta =\sqrt{\left( V_{f}+B_{0}\right)
^{2}-4D_{0}\rho }.
\end{equation*}
Hence we obtain
\begin{equation}  \label{36}
\theta \left( \xi \right) =\frac{\Delta }{2\left( V_{f}+B_{0}\right) }\tanh %
\left[ \frac{\Delta }{2D_{0}}\left( \xi -\xi _{0}\right) \right] -\frac{1}{2}%
,\frac{\rho D_{0}}{\left( V_{f}+B_{0}\right) ^{2}}\neq \frac{1}{4},
\end{equation}
The solution given by Eq.~(\ref{36}) covers both cases $\left(
V_{f}+B_{0}\right) ^{2}-4D_{0}\rho >0$ and $\left( V_{f}+B_{0}\right)
^{2}-4D_{0}\rho <0$, respectively. In the case $\left( V_{f}+B_{0}\right)
^{2}-4D_{0}\rho <0$ with the use of the identity $\tanh (ix)=i\tan(x)$, the
solutions can be expressed in terms of trigonometric functions.  For the
case $k=1/4$ we obtain
\begin{equation*}
\left( \xi -\xi _{0}\right) \frac{V_{f}+B_{0}}{D_{0}}=\int \frac{d\theta }{%
\theta ^{2}+\theta +k}=4\int \frac{d\theta }{\left( 2\theta +1\right) ^{2}},%
\frac{\rho D_{0}}{\left( V_{f}+B_{0}\right) ^{2}}=\frac{1}{4},
\end{equation*}
\begin{equation}  \label{38}
\theta (\xi )=-\frac{1}{2}-\frac{D_{0}}{V_{f}+B_{0}}\frac{1}{\xi -\xi _{0}},%
\frac{\rho D_{0}}{\left( V_{f}+B_{0}\right) ^{2}}=\frac{1}{4},
\end{equation}

After substituting the expressions of $\theta (\xi )$ given by Eqs.~(\ref{36}%
) and (\ref{38}) into Eqs.~(\ref{33}) and (\ref{34}) we obtain the general
solution of Eq.~(\ref{ex1}) as
\begin{eqnarray}
u(\xi ) &=&\frac{\left( V_{f}+B_{0}\right) ^{2}}{CD_{0}\rho }e^{-\left(
V_{f}+B_{0}\right) \left( \xi -\xi _{0}\right) /2D_{0}}\Bigg\{\cosh \left[
\frac{\Delta }{2D_{0}}\left( \xi -\xi _{0}\right) \right] +  \notag \\
&&\frac{\Delta }{V_{f}+B_{0}}\sinh \left[ \frac{\Delta }{2D_{0}}\left( \xi
-\xi _{0}\right) \right] \Bigg\},\frac{\rho D_{0}}{\left( V_{f}+B_{0}\right)
^{2}}\neq \frac{1}{4},  \label{dd}
\end{eqnarray}%
\begin{equation}
u(\xi )=\frac{(V_{f}+B_{0})^{2}}{4CD_{0}^{2}\rho }\left( \xi -\xi _{0}+\frac{%
2D_{0}}{V_{f}+B_{0}}\right) e^{-\frac{\left( V_{f}+B_{0}\right) \left( \xi
-\xi _{0}\right) }{2D_{0}}},\frac{\rho D_{0}}{\left( V_{f}+B_{0}\right) ^{2}}%
=\frac{1}{4},  \label{dd1}
\end{equation}%
where $C$ and $\xi _{0}$ are the arbitrary constants of integration to be
determined from the initial conditions.

Note that the general solution of Eq.~(\ref{ex1}), as given by Eqs.~(\ref{dd}%
) and (\ref{dd1}) can be obtained directly from the initial Eq.~(\ref{ex1}),
which is a linear second order homogeneous differential equation, and whose
solution is immediate.

\subsection{Travelling wave solutions for $D(u)\propto
\left(1-u/u_{max}\right)^{-1}$, $B(u)=\mathrm{constant}$, $Q(u)\propto
u\left(1-u/u_{max}\right)$}

As a second example of application of \textbf{Theorem 1} we consider that
the diffusion, convection and the reaction functions are given by
\begin{equation*}  \label{42}
D(u)=\frac{D_0}{1-u/u_{max}},B(u)=B_0,Q(u)=\rho u\left(1-\frac{u}{u_{max}}%
\right),D_0,B_0,\rho ,u_{max} =\mathrm{constant}.
\end{equation*}

The travelling wave equation for the reaction-convection-diffusion system
with these forms of $D$, $B$ and $Q$ is given by
\begin{equation}
\frac{d^{2}u}{d\xi ^{2}}+\frac{1}{u_{max}\left( 1-u/u_{max}\right) }\left(
\frac{du}{d\xi }\right) ^{2}+\frac{V_{f}+B_{0}}{D_{0}}\left( 1-\frac{u}{%
u_{max}}\right) \frac{du}{d\xi }+\frac{\rho }{D_{0}}u\left( 1-\frac{u}{%
u_{max}}\right) ^{2}=0.  \label{33}
\end{equation}

Eq.~(\ref{33}) represents the travelling wave form for a generalized Fisher
equation given by
\begin{equation*}
\frac{\partial u(x,t)}{\partial t}=\frac{\partial }{\partial x}\left[ \frac{%
D_{0}}{1-u(x,t)/u_{max}}\frac{\partial u(x,t)}{\partial x}\right] +B_0\frac{%
\partial u(x,t)}{\partial x}+\rho u(x,t)\left[ 1-\frac{u(x,t)}{u_{max}}%
\right] .
\end{equation*}

Since the functions $D(u)$, $B(u)$ and $Q(u)$ satisfy the integrability
condition given by Eq.~(\ref{cond}), with
\begin{equation*}
k=\frac{D_0\rho }{\left(V_f+B_0\right)^2},
\end{equation*}
the general solution of Eq.~(\ref{33}) can be obtained in an exact
parametric form.

\subsubsection{The case $k=1/4$}

We consider first the case $k=1/4$. Then the general solution of Eq.~(\ref%
{33}) is given in parametric form as
\begin{equation*}
\xi (\theta )-\xi _{0}=\pm 2CD_{0}u_{max}\int_{\theta _{0}}^{\theta }\frac{%
\,d\psi }{\left( 1+2\psi \right) \left[ C\sqrt{D_{0}\rho }u_{max}(1+2\psi
)\mp 2\psi e^{\frac{1}{1+2\psi }}\right] },
\end{equation*}
\begin{equation*}
u(\theta )=\pm \frac{2}{C\sqrt{\rho D_0}}\frac{\theta }{1+2\theta}\exp
\left( \frac{1}{1+2\theta }\right) ,V_{f}+B_{0}=\pm 2\sqrt{D_{0}\rho },
\end{equation*}
where $C$ is an arbitrary constant of integration.

\subsubsection{The case $k\neq 1/4$}

For $k\neq 1/4$, we obtain
\begin{eqnarray*}
\xi (\theta ,k)-\xi _{0}&=&\frac{D_{0}Cu_{max}}{V_{f}+B_{0}}\int_{\theta
_{0}}^{\theta }{\left[ Cu_{max}\left( \psi ^{2}+\psi +k\right) -\frac{\psi
\sqrt{\psi ^{2}+\psi +k}}{e^{\frac{\tan ^{-1}\left( \frac{2k\psi +1}{\sqrt{%
4k-1}}\right) }{\sqrt{4k-1}}}}\right] ^{-1}d\psi },  \notag \\
&&k=\frac{\rho D_{0}}{\left( V_{f}+B_{0}\right) ^{2}}>1/4,
\end{eqnarray*}
\begin{equation*}
u(\theta ,k)=\frac{V_{f}+B_{0}}{C\rho D_{0}}\frac{\theta }{\sqrt{\theta
^{2}+\theta +k}}\exp \left( -\frac{1}{\sqrt{4k-1}}\tan ^{-1}\frac{1+2k\theta
}{\sqrt{4k-1}}\right) ,k=\frac{\rho D_{0}}{\left( V_{f}+B_{0}\right) ^{2}}%
>1/4,
\end{equation*}
\begin{eqnarray*}
\xi (\theta ,k)-\xi _{0} &=&\frac{D_{0}Cu_{max}}{V_{f}+B_{0}}\int_{\theta
_{0}}^{\theta }{\left[ Cu_{max}\left( \psi ^{2}+\psi +k\right) -\frac{\psi
\sqrt{\psi ^{2}+\psi +k}}{e^{\frac{\mathrm{tanh}^{-1}\left( \frac{2k\psi +1}{%
\sqrt{1-4k}}\right) }{\sqrt{1-4k}}}}\right] ^{-1}d\psi },  \notag \\
&&k=\frac{\rho D_{0}}{\left( V_{f}+B_{0}\right) ^{2}}<1/4,
\end{eqnarray*}
\begin{equation*}
u(\theta ,k)=\frac{V_{f}+B_{0}}{C\rho D_{0}}\frac{\theta }{\sqrt{\theta
^{2}+\theta +k}}\exp \left( \frac{1}{\sqrt{1-4k}}\tanh ^{-1}\frac{1+2k\theta
}{\sqrt{1-4k}}\right) ,k=\frac{\rho D_{0}}{\left( V_{f}+B_{0}\right) ^{2}}%
<1/4,
\end{equation*}
where the constant $k$ is determined by the model parameters $V_{f}$, $D_{0}$%
, $B_{0}$ and $\rho $, respectively.

\subsection{Travelling wave solutions of the reaction--convection--diffusion
equation with power law diffusion and reaction functions}

Exact travelling wave solutions of more general
reaction-convection-diffusion equations of the form
\begin{equation}
\frac{\partial u(x,t)}{\partial t}=\frac{\partial }{\partial x}\left\{ \frac{%
D_{0}}{\left[ 1-u(x,t)/u_{max}\right] ^{\alpha }}\frac{\partial u(x,t)}{%
\partial x}\right\} +B_{0}\frac{\partial u(x,t)}{\partial x}+\rho u(x,t)%
\left[ 1-\frac{u(x,t)}{u_{max}}\right] ^{\alpha },  \label{mod0}
\end{equation}%
corresponding to
\begin{equation}  \label{42n}
D(u)=\frac{D_0}{\left(1-u/u_{max}\right)^{\alpha }},B(u)=B_0,Q(u)=\rho
u\left(1-\frac{u}{u_{max}}\right)^{\alpha },D_0,B_0,\rho ,u_{max}, \alpha =%
\mathrm{constant},
\end{equation}
are given, in a parametric form, by the following equations,
\begin{equation}
\xi (\theta ,k)-\xi _{0}\left( \theta _{0},k\right) =\frac{D_{0}}{V_{f}+B_{0}%
}\int_{\theta _{0}}^{\theta }\frac{{d\psi }}{\left[
1-C^{-1}u_{max}^{-1}e^{F(\psi ,k)}\right] ^{\alpha }\left( \psi ^{2}+\psi
+k\right) },  \label{mod1}
\end{equation}
and
\begin{equation}
u(\theta ,k)=\frac{V_f+B_0}{CD_0\rho }e^{F(\theta ,k)},  \label{mod2}
\end{equation}%
respectively, with $k=\rho D_{0}/\left( V_{f}+B_{0}\right) ^{2}$. In this
mathematical model the diffusion $D(u)$ and the reaction function $Q(u)$
have a power-law type dependence on the function $u$. Due to this specific
dependence in the following we will call Eq.~(\ref{mod0}) as \textit{%
reaction--convection--diffusion equation with power law diffusion and
reaction functions}.

\subsection{Travelling wave solutions with $B(u)=\pm B_0u/\protect\sqrt{1+%
\frac{kB_0^2}{\protect\rho D_{0}}u^{2}}-V_{f}$, $D(u)\propto
\left(1-u/u_{max}\right)^{-\protect\alpha}$, $Q(u)\propto
u\left(1-u/u_{max}\right)^{\protect\alpha }$}

In the following we assume again that the diffusion function $D(u)$ and the
reaction function $Q(u)$ have the forms given by Eq.~(\ref{42n}). Next, we
determine the most general convection function $B(u)$ that leads to exact
travelling wave solutions of Eq.~(\ref{1}) for these choices of the
diffusion and reaction functions.

From the Chiellini lemma it follows that in order for the corresponding Abel
equation be exactly integrable, the convection function must satisfy the
differential condition, which immediately follows from Eq.~(\ref{cond}),
\begin{equation*}
\frac{d}{du}\left[ \frac{D_{0}\rho u}{V_{f}+B(u)}\right] =k\left[ V_{f}+B(u)%
\right],
\end{equation*}
a condition which reduces to a first order differential equation for $B(u)$,
\begin{equation*}
u\frac{d}{du}\left[ V_{f}+B(u)\right] =V_{f}+B(u)-\frac{k}{\rho D_{0}}\left[
V_{f}+B(u)\right] ^{3},
\end{equation*}
with the general solution given by
\begin{equation*}
B(u)=\pm \frac{B_0u}{\sqrt{1+\frac{kB_0^2}{\rho D_{0}}u^{2}}}-V_{f},
\end{equation*}
where $B_{0}$ is an arbitrary constant of integration.

The general reaction-convection-diffusion equation takes the form
\begin{eqnarray*}
\frac{\partial u(x,t)}{\partial t} &=&\frac{\partial }{\partial x}\left\{
\frac{D_{0}}{\left[ 1-u(x,t)/u_{max}\right] ^{\alpha }}\frac{\partial u(x,t)%
}{\partial x}\right\} +\left( \pm \frac{B_{0}u}{\sqrt{1+\frac{kB_{0}^{2}}{%
\rho D_{0}}u^{2}}}-V_{f}\right) \frac{\partial u(x,t)}{\partial x}+  \notag
\\
&&\rho u(x,t)\left[ 1-\frac{u(x,t)}{u_{max}}\right] ^{\alpha },
\end{eqnarray*}%
and it has exact travelling wave solutions, given in a parametric form as
\begin{equation*}
\xi \left( \theta ,k\right) -\xi _{0}\left( \theta _{0},k\right) =\pm \sqrt{%
\frac{kD_{0}}{\rho }}\int_{\theta _{0}}^{\theta }\frac{\left[ 1\mp \left(
\sqrt{\rho D_{0}/kB_{0}^{2}u_{max }^{2}}\right) \sqrt{\left(
B_{0}/CD_{0}\rho \right) ^{2}e^{2F\left( \psi ,k\right) }-1}\right]
^{-\alpha }d\psi }{\sqrt{1-\left( CD_{0}\rho /B_{0}\right) ^{2}e^{-2F\left(
\psi ,k\right) }}\left( \psi ^{2}+\psi +k\right) },
\end{equation*}
\begin{equation*}
u(\theta ,k)=\pm \sqrt{\frac{\rho D_{0}}{kB_{0}^{2}}}\sqrt{\left( \frac{B_{0}%
}{CD_{0}\rho }\right) ^{2}e^{2F(\theta ,k)}-1}.
\end{equation*}

\section{Exact travelling wave solutions obtained from the Lemke
integrability condition of the Abel equation}

\label{sect3}

In the present Section we consider an alternative method to obtain exact
solutions of the Abel equation, which is based on the so-called Lemke
transformation. In this method through a series of successive
transformations the Abel equation is reduced to a second order non-linear
differential equation. This equation can be exactly integrated in a number
of cases, thus leading to some exact classes of travelling wave solutions.

\subsection{Transformation of the Abel equation into a second order
differential equation}

By introducing a new independent variable $\tau $ through the Lemke
transformation \cite{Lemke, kamke}, defined as,
\begin{equation}
\tau =\int {f(u)du},u=u(\tau ),  \label{trans}
\end{equation}%
the Abel Eq.~(\ref{5}) takes the form
\begin{equation}
\frac{dw(\tau )}{d\tau }=w^{2}(\tau )+\frac{g(\tau )}{f(\tau )}w^{3}(\tau ).
\label{35}
\end{equation}%
Eq.~(\ref{35}) must be integrated with the initial condition $w\left( \tau
_{0}\right) =w_{0}$, where $\tau _{0}=\left. \int {f(u)du}\right\vert
_{u=u_{0}}$. Next we introduce the function $\eta (\tau )$ defined as
\begin{equation}
w(\tau )=-\frac{1}{\eta (\tau )}\frac{d\eta (\tau )}{d\tau }.  \label{sa}
\end{equation}

In terms of $\eta $, Eq.~(\ref{35}) becomes
\begin{equation}  \label{fin0}
\frac{d^2\eta (\tau )}{d\tau ^2}=\frac{g(\tau )}{f(\tau )}\frac{1}{\eta
^2(\tau )}\left[\frac{d\eta (\tau )}{d\tau }\right]^3.
\end{equation}
By taking into account the mathematical identity
\begin{equation}
\frac{d^2y}{dx^2}=-\frac{d^2x}{dy^2}\left(\frac{dy}{dx}\right)^{3},
\end{equation}
Eq.~(\ref{fin0}) takes the form
\begin{equation}  \label{fin}
\eta ^2 \frac{d^2\tau }{d\eta ^2}+\frac{D(u(\tau))Q(u(\tau) )}{V_f+B(u(\tau
))}=0.
\end{equation}

Eq.~(\ref{fin}) must be integrated with the initial conditions $\tau \left(
\eta _{0}\right) =\tau _{0}$ and $\left. \left( d\tau /d\eta \right)
\right\vert _{\eta =\eta _{0}}=\tau _{0}^{\prime }$, respectively.

\subsection{The Lemke integrability condition of the general reaction--convection--diffusion equation}

The \textit{Lemke integrability condition} for the general
reaction--diffusion--convection Eq.~(\ref{1}) can be formulated as the
following

\textbf{Theorem 2}. \textit{If a solution $\eta =\eta (\tau )$ of Eq.~(\ref%
{fin}) is known, then the general reaction-convection-diffusion Eq.~(\ref{1}%
) admits travelling wave solutions, given in a parametric form as
\begin{equation}
u=u(\eta ),\xi (\eta )-\xi _{0}\left( \eta _{0}\right) =-\int_{\eta
_{0}}^{\eta }\eta {^{-1}\frac{D(u(\eta ))}{V_{f}+B(u(\eta ))}d\eta },
\label{67}
\end{equation}%
where $\xi _{0}\left( \eta _{0}\right) $ is an arbitrary constant of
integration. If the solution of Eq.~(\ref{fin}) is obtained in a parametric
form, with the parameter $\theta $, then the general travelling wave
solution of the reaction-convection-diffusion Eq.~(\ref{1}) is obtained as
\begin{equation}
\tau =\tau (\theta ),\eta =\eta (\theta ),u=u[\tau (\theta )],
\end{equation}%
\begin{equation}
\xi \left( \theta \right) -\xi _{0}\left( \theta _{0}\right) =-\int_{\theta
_{0}}^{\theta }\frac{1}{\eta \left( \theta \right) }\frac{D(u(\tau (\theta
)))}{V_{f}+B(u(\tau (\theta )))}\frac{d\eta \left( \theta \right) }{d\theta }%
d\theta .  \label{69}
\end{equation}%
}

\textbf{Proof.} Assume that a solution $\eta =\eta (\tau )$ of Eq.~(\ref{fin}%
) is known. The Lemke transformation Eq.~(\ref{trans}) takes the form
\begin{equation}
\tau =V_{f}u+\int {B(u)du},
\end{equation}%
giving
\begin{equation}
\frac{du}{d\tau }=\frac{1}{V_{f}+B(u)}.
\end{equation}%
In view of Eqs.~(\ref{k1},\ref{kkk},\ref{sa}), then we obtain the relations
\begin{equation}
\frac{du}{d\xi }=\frac{du}{d\tau }\frac{d\tau }{d\xi }=\frac{1}{D(u(\tau
))w(u(\tau ))}=-\frac{\eta (\tau )}{D(u(\tau ))\left( d\eta (\tau )/d\tau
\right) },
\end{equation}%
leading to the result
\begin{equation}
d\xi =-\frac{\eta
^{-1}D(u(\eta ))}{V_{f}+B(u(\eta ))}d\eta ,
\end{equation}
which by integration gives the second of Eqs.~(\ref{67}).

Now assume that the general solution of Eq.~(\ref{fin}) is known in a
parametric form, so that we get
\begin{equation}
\tau =\tau \left( \theta \right) ,\eta =\eta \left( \theta \right) ,u=u\left[
\tau \left( \theta \right) \right] .
\end{equation}%
Therefore we obtain first
\begin{equation}
\frac{du}{d\xi }=\frac{du}{d\tau }\frac{d\tau }{d\theta }\frac{d\theta }{%
d\xi }=\frac{1}{D(\theta )w(\theta )}=-\frac{\eta (\theta )}{D(\theta
)\left( d\eta /d\theta \right) \left( d\theta /d\tau \right) },
\end{equation}%
giving
\begin{equation}
d\xi =-\frac{1}{\eta (\theta )}\frac{D(u(\tau (\theta )))}{V_{f}+B(u(\tau
(\theta )))}\frac{d\eta (\theta )}{d\theta }d\theta ,
\end{equation}%
which by integration gives Eq.~(\ref{69}). This concludes the proof of
\textbf{Theorem 2}.

Depending on the functional form of the product of the diffusion and
reaction functions $D(c)$ and $Q(c)$, Eq.~(\ref{fin}) can be integrated
exactly in a number of cases, thus leading, with the help of \textbf{Theorem
2}, to exact travelling wave solutions of the general
reaction-convection-diffusion Eq.~(\ref{1}), which we present in the
following.

\subsection{Travelling wave solutions for $D(u(\protect\tau))Q(u(\protect\tau%
))=\mathrm{constant}$, $B(u(\protect\tau))=\mathrm{constant}$}

In the case the product of the diffusion and reaction functions $D(u(\tau
))Q(u(\tau ))$ is a constant $\alpha $,
\begin{equation}
D(u(\tau ))Q(u(\tau ))=\alpha ,  \label{d1}
\end{equation}%
and the convection function $B(u(\tau ))$ is a constant,
\begin{equation}\label{d2n}
B(u(\tau ))=B_{0}=\mathrm{constant},
\end{equation}%
Eq.~(\ref{fin}) takes the form
\begin{equation*}
\eta ^{2}\frac{d^{2}\tau }{d\eta ^{2}}+\beta =0,
\end{equation*}%
where $\beta =\alpha /\left( V_{f}+B_{0}\right) =\mathrm{constant}$, and it
has the general solution
\begin{equation*}
\tau (\eta )=C_{1}+C_{2}\eta +\beta \ln \left\vert \eta \right\vert ,
\end{equation*}%
where $C_{1}$ and $C_{2}$ are arbitrary constants of integration. Since $%
f(u)=V_{f}+B(u)=V_{f}+B_{0}$, from Eq.~(\ref{trans}) it follows that
\begin{equation*}
u(\tau )=\frac{\tau }{V_{f}+B_{0}}.
\end{equation*}%
For $w$ we obtain
\begin{equation}
w\left( \eta \right) =-\frac{1}{\eta (\tau )}\frac{d\eta (\tau )}{d\tau }=-%
\frac{1}{\eta (\tau )}\frac{1}{\frac{d\tau (\eta )}{d\eta }}=-\frac{1}{\beta
+C_{2}\eta }.  \label{ww}
\end{equation}

Eq.~(\ref{ww}) identically satisfies Eq.~(\ref{5}), and therefore we can
take the arbitrary integration constant $C_{1}$ as zero, $C_{1}=0$. Using
the transformations
\begin{equation*}
v\left( \eta \right) =D\left( \eta \right) w\left( \eta \right) =-\frac{%
D(\eta )}{\left( \beta +C_{2}\eta \right) },
\end{equation*}
\begin{equation*}
\frac{du}{d\xi }=\left( \frac{du}{d\eta }\right) \left( \frac{d\eta }{d\xi }%
\right) =\frac{1}{v\left( \eta \right) }=-\frac{\left( \beta +C_{2}\eta
\right) }{D(\eta )},
\end{equation*}%
and with the help of Eq.~(\ref{ww}), we obtain the following expressions for
$u$ and $\xi $, giving, the general solution of the general
reaction-convection-diffusion Eq.~(\ref{1}), with the coefficients $D(u)$
and $Q(u)$ satisfying the condition (\ref{d1}), and $B(u)=\mathrm{constant}$%
, as,
\begin{equation}
u(\eta )=\frac{C_{2}}{V_{f}+B_{0}}\eta +\frac{\alpha }{\left(
V_{f}+B_{0}\right) ^{2}}\ln |\eta |,  \label{c1n}
\end{equation}%
\begin{eqnarray}
\xi (\eta )-\xi _{0}\left( \eta _{0}\right) &=&-\frac{1}{V_f+B_0}\int_{\eta
_{0}}^{\eta }\psi ^{-1}{D\left[ \frac{C_{2}}{V_{f}+B_{0}}\psi +\frac{\alpha
}{\left( V_{f}+B_{0}\right) ^{2}}\ln |\psi |\right] d\psi },  \label{xi1}
\end{eqnarray}%
where $\xi _{0}\left( \eta _{0}\right) $ is an arbitrary constant of
integration, and we have taken $\eta $ as a parameter. Eqs.~(\ref{c1n}) and (%
\ref{xi1}) give the general solution of the travelling wave form of the
general reaction-convection-diffusion Eq.~(\ref{3}), with diffusion,
convection and reaction functions satisfying the conditions (\ref{d1}) and (%
\ref{d2n}). In order to determine the integration constants $\eta _{0}$ and $%
C_{2}$, we use the initial conditions that give
\begin{equation*}
u(\eta _{0})=\frac{C_{2}}{V_{f}+B_{0}}\eta _{0}+\frac{\alpha }{\left(
V_{f}+B_{0}\right) ^{2}}\ln |\eta _{0}|=u_{0},
\end{equation*}%
and
\begin{equation*}
\left. \frac{du}{d\xi }\right\vert _{\xi =0}=\left. \frac{du}{d\eta }\frac{%
d\eta }{d\xi }\right\vert _{\eta =\eta _{0}}=-\left( C_{2}+\frac{\alpha }{%
V_{f}+B_{0}}\frac{1}{\eta _{0}}\right) \frac{\eta _{0}}{D\left( u\left( \eta
_{0}\right) \right) }=u_{0}^{\prime },
\end{equation*}%
where, with the use of Eq.~(\ref{xi1}), we have $d\eta /d\xi =1/d\xi /d\eta
=-\left(V_f+B_0\right)\eta /D(\eta )$, giving
\begin{equation*}
C_{2}=-\frac{1}{\eta _{0}}\left[ \frac{\alpha }{V_{f}+B_{0}}+D\left( u\left(
\eta _{0}\right) \right) u_{0}^{\prime }\right] ,
\end{equation*}%
and
\begin{equation}
\ln \eta _{0}=\frac{\left( V_{f}+B_{0}\right) ^{2}}{\alpha }u_{0}+\frac{%
\left( V_{f}+B_{0}\right) u_{0}^{\prime }}{\alpha }D\left( u\left( \eta
_{0}\right) \right) +1,  \label{61n}
\end{equation}%
respectively. Eqs.~(\ref{61n}) determines the value of $\eta _{0}$ as a
function of the initial conditions $\left( u_{0},u_{0}^{\prime }\right) $
and of the free parameters $\left\{ \alpha ,B_{0},V_{f},D\left( u\left( \eta
_{0}\right) \right) \right\} $.

A similar travelling wave solution of the reaction-convection-diffusion Eq.~(%
\ref{1}) is obtained if the functions $D$,$B$ and $Q$ satisfy the general
relation
\begin{equation*}
\frac{D(u(\tau ))Q(u(\tau ))}{V_{f}+B(u(\tau ))}=\beta =\mathrm{constant}.
\end{equation*}

\subsection{Travelling wave solutions with a linear dependence of $D(u(%
\protect\tau))Q(u(\protect\tau))$ and constant $B(u(\protect\tau))$}

Next, in order to obtain another exact general solution of Eq.~(\ref{fin}),
we assume that the diffusion function $D\left( u(\tau )\right) $, the
reaction function $Q\left( u(\tau)\right) $ and the convection function $%
B(u(\tau))$ satisfy the conditions
\begin{equation}
D\left( u(\tau)\right) Q\left( u(\tau)\right) =\beta \tau +\alpha ,
B(u(\tau))=B_0=\mathrm{constant},  \label{dc}
\end{equation}%
where $\alpha $, $\beta $ and $B_0$ are arbitrary constants. By inserting
Eq.~(\ref{dc}) into Eq.~(\ref{fin}), the latter equation becomes
\begin{equation}
\eta ^{2}\frac{d^{2}\tau }{d\eta ^{2}}+\frac{\beta \tau+\alpha }{V_{f}+B_0}%
=0.  \label{dc1}
\end{equation}

Eq.~(\ref{dc1}) can be integrated to yield
\begin{equation*}
\tau (\eta )=c_{1}\eta ^{m_{+}}+c_{2}\eta ^{m_{-}}-\frac{\alpha }{\beta },
\end{equation*}%
where $c_{1}$ and $c_{2}$ are the arbitrary constants of integration, and
\begin{equation*}
2m_{\pm }=1\pm \sqrt{1-\frac{4\beta }{V_{f}+B_{0}}}.
\end{equation*}

The function $w(\eta)$ is given by
\begin{equation*}
w(\eta )=-\frac{1}{\eta }\frac{1}{\left(d\tau /d\eta\right)}=-\frac{1}{%
c_1m_+\eta ^{m_+}+c_2m_{-}\eta ^{m_{-}}}.
\end{equation*}

The function $v\left( \eta \right) $ is obtained as
\begin{equation*}
v\left( \eta \right) =-\frac{D\left( u(\eta )\right) }{ c_{1}m_{+}\eta
^{m_{+}}+c_{2}m_{-}\eta ^{m_{-}} }.
\end{equation*}

Therefore the general solution of a reaction--convection--diffusion equation
with coefficients $D(u)$, $B(u)$ and $Q(u)$ satisfying the conditions (\ref%
{dc}) is given, in a parametric form, by
\begin{equation*}
u(\eta )=\frac{1}{V_{f}+B_{0}}\left( c_{1}\eta ^{m_{+}}+c_{2}\eta ^{m_{-}}-%
\frac{\alpha }{\beta }\right) ,
\end{equation*}%
\begin{equation*}
\xi (\eta )-\xi _{0}\left( \eta _{0}\right) =-\frac{1}{V_{f}+B_{0}}%
\int_{\eta _{0}}^{\eta }\frac{c_{1}m_{+}\psi ^{m_{+}-1}+c_{2}m_{-}\psi
^{m_{-}-1}}{c_{1}m_{+}\psi ^{m_{+}}+c_{2}m_{-}\psi ^{m_{-}}}D\left[ \frac{1}{%
V_{f}+B_{0}}\left( c_{1}\psi ^{m_{+}}+c_{2}\psi ^{m_{-}}-\frac{\alpha }{%
\beta }\right) \right] {d\psi },  \label{xi2}
\end{equation*}
where $\xi _{0}\left( \eta _{0}\right) $ is an arbitrary constant of
integration, and we have taken $\eta $ as a parameter.

The numerical values of the integration constants $c_{1}$ and $c_{2}$ are
determined from the initial conditions as
\begin{equation*}
c_{1}=\frac{\eta _{0}^{-m_{+}}\left\{ m_{-}\left[ \alpha +\beta
u_{0}(B_{0}+V_{f})\right] -\beta \eta _{0}u_{0}^{\prime }\left( \eta
_{0}\right) (B_{0}+V_{f})\right\} }{\beta (m_{-}-m_{+})},
\end{equation*}%
and
\begin{equation*}
c_{2}=\frac{\eta _{0}^{-m_{-}}\left\{ \beta \eta _{0}u_{0}^{\prime }\left(
\eta _{0}\right) (B_{0}+V_{f})-m_{+}\left[ \alpha +\beta u_{0}(B_{0}+V_{f})%
\right] \right\} }{\beta (m_{-}-m_{+})},
\end{equation*}%
respectively, where $u_{0}=u\left( \eta _{0}\right) $ and $u_{0}^{\prime
}\left( \eta _{0}\right) =\left. \left( du/d\eta \right) \right\vert _{\eta
=\eta _{0}}$. By taking into account the relation
\begin{equation*}
\left. \frac{du}{d\xi }\right\vert _{\xi =0}=u_{0}^{\prime }=\left. \frac{du%
}{d\eta }\right\vert _{\eta =\eta _{0}}\left. \frac{d\eta }{d\xi }%
\right\vert _{\eta =\eta _{0}}=-u_{0}^{\prime }\left( \eta _{0}\right) \frac{%
\left( V_{f}+B_{0}\right) }{D\left( u\left( \eta _{0}\right) \right) }\frac{%
c_{1}m_{+}\eta _{0}^{m_{+}}+c_{2}m_{-}\eta _{0}^{m_{-}}}{c_{1}m_{+}\eta
_{0}^{m_{+}-1}+c_{2}m_{-}\eta _{0}^{m_{-}-1}},
\end{equation*}%
we obtain
\begin{equation*}
u_{0}^{\prime }\left( \eta _{0}\right) =-\frac{D\left( u\left( \eta
_{0}\right) \right) }{\left( V_{f}+B_{0}\right) }\left( \frac{c_{1}m_{+}\eta
_{0}^{m_{+}-1}+c_{2}m_{-}\eta _{0}^{m_{-}-1}}{c_{1}m_{+}\eta
_{0}^{m_{+}}+c_{2}m_{-}\eta _{0}^{m_{-}}}\right) u_{0}^{\prime }.
\end{equation*}

As for the initial value $\eta _{0}$ of the parameter $\eta $, it can be
determined from the condition given by Eq.~(\ref{dc}), once the explicit
functional forms of the functions $D\left( u(\tau )\right) $ and $Q\left(
u(\tau )\right) $ are known.

If the arbitrary constant $\alpha $ vanishes, then we obtain the condition $%
D(u(\tau))Q(u(\tau))=\beta \tau$, given by Eq.~(\ref{cond}), and thus we
regain the solution obtained in the previous Section by using the Chiellini
lemma.

\subsection{Travelling wave solutions with $D(u(\protect\tau))Q(u(\protect%
\tau))\propto 1/\protect\tau$, $B(u)=\mathrm{constant}$}

If the diffusion, convection and reaction functions $D(u(\tau ))$, $Q(u(\tau
))$, $B(u(\tau ))$ satisfy the relations
\begin{equation}
D(u(\tau ))Q(u(\tau ))=\frac{K\left( V_{f}+B_{0}\right) }{\tau },B(u(\tau
))=B_{0}=\mathrm{constant},  \label{d3}
\end{equation}%
where $K$ is an arbitrary constant, Eq.~(\ref{fin}) takes the form of an
Emden-Fowler equation \cite{Pol},
\begin{equation}
\eta ^{2}\frac{d^{2}\tau }{d\eta ^{2}}+K\frac{1}{\tau }=0.  \label{50}
\end{equation}%
The general solution of Eq.~(\ref{50}) can be obtained in parametric form,
with $\theta $ taken as parameter, as \cite{Pol},
\begin{equation*}
\eta (\theta )=k_{1}\left[ \left( \sqrt{\pi }/2\right) \mathrm{erf}(\theta
)+k_{2}\right] ^{-1},  \label{c3}
\end{equation*}%
\begin{equation*}
\tau (\theta )=\sqrt{\frac{K}{2}}\frac{d}{d\theta }\ln \left[ \left( \sqrt{%
\pi }/2\right) \mathrm{erf}(\theta )+k_{2}\right] ,
\end{equation*}%
where $k_{1}$ and $k_{2}$ are arbitrary constants of integration, and $%
\mathrm{erf}(z)=\left( 2/\sqrt{\pi }\right) \int_{0}^{z}{e^{-t^{2}}dt}$ is
the error function, representing the integral of the Gaussian distribution.
Since by definition $\tau =\int{f(u)du}=\left(V_f+B_0\right)u$, we obtain $%
u(\theta )=\tau (\theta )/\left(V_f+B_0\right)$, and therefore the general
solution of the reaction-convection-diffusion equation with coefficients
satisfying Eqs.~(\ref{d3}) is obtained in a parametric form as
\begin{equation*}
u(\theta )=\frac{1}{V_{f}+B_{0}}\sqrt{\frac{K}{2}}\frac{d}{d\theta }\ln %
\left[ \left( \sqrt{\pi }/2\right) \mathrm{erf}(\theta )+k_{2}\right] ,
\end{equation*}%
\begin{equation*}
\xi (\theta )-\xi _{0}\left( \theta _{0}\right) =\frac{2}{V_{f}+B_{0}}%
\int_{\theta _{0}}^{\theta }{\frac{e^{-\psi ^{2}}D(u(\psi ))}{2k_{2}+\sqrt{%
\pi }\text{erf}(\psi )}d\psi }.  \label{xi3}
\end{equation*}

\subsection{Travelling waves solutions with a power law dependence of $D(u(%
\protect\tau))Q(u(\protect\tau))$ and constant $B(u)$}

If $D(u(\tau ))$, $Q(u(\tau ))$ and $B(u(\tau ))$ satisfy the relations
\begin{equation}
\frac{1}{V_{f}+B_{0}}D(u(\tau ))Q(u(\tau ))=\frac{2(m+1)}{(m+3)^{2}}\tau
-A\tau ^{m},m\neq -1,m\neq -3,  \label{d4}
\end{equation}%
where $m$ and $A$ are arbitrary constants, and $B(u(\tau ))=B_{0}=\mathrm{%
constant}$, then the basic equation determining the travelling wave
solutions of the corresponding general reaction-convection-diffusion
equation is
\begin{equation}
\eta ^{2}\frac{d^{2}\tau }{d\eta ^{2}}+\frac{2(m+1)}{(m+3)^{2}}\tau -A\tau
^{m}=0.  \label{mods0}
\end{equation}%
Eq.~(\ref{mods0}) has the exact parametric solution given by \cite{Pol}
\begin{equation*}
\eta \left( \theta \right) =l_{1}\left( \int^{\theta }{\frac{d\psi }{\sqrt{%
1\pm \psi ^{m+1}}}}+l_{2}\right) ^{\left( m+3\right) /(m-1)},m\neq -1,m\neq
-3,m\neq 1,
\end{equation*}%
\begin{equation}
\tau \left( \theta \right) =b\theta \left( \int^{\theta }\frac{d\psi }{\sqrt{%
1\pm \psi ^{m+1}}}+l_{2}\right) ^{2/(m-1)},m\neq -1,m\neq -3,m\neq 1,
\label{c4}
\end{equation}%
where
\begin{equation*}
A=\pm \frac{(m+1)(m-1)^{2}}{2(m+3)^{2}}b^{1-m},
\end{equation*}%
and $l_{1}$ and $l_{2}$ are the arbitrary constants of integration and $b$
is an arbitrary constant. Then the general solution of the corresponding
reaction-convection-diffusion equation is obtained in a parametric form as
\begin{equation*}
u_{\pm }(\theta )=\frac{b}{V_{f}+B_{0}}\theta \left( \int_{\theta
_{0}}^{\theta }\frac{d\psi }{\sqrt{1\pm \psi ^{m+1}}}+l_{2}\right)
^{2/(m-1)},m\neq -1,m\neq -3,m\neq 1,
\end{equation*}%
\begin{equation*}
\xi _{\pm }(\theta )-\xi _{0\pm }\left( \theta _{0}\right) =-\frac{m+3}{%
(m-1)\left( V_{f}+B_{0}\right) }\int_{\theta _{0}}^{\theta }{\frac{D(u(\psi
))d\psi }{\sqrt{1\pm \psi ^{m+1}}\left[ l_{2}+\psi \,_{2}F_{1}\left( \frac{1%
}{2},\frac{1}{m+1};1+\frac{1}{m+1};-\psi ^{m+1}\right) \right] }},
\end{equation*}%
where $\xi _{0\pm }$ are the arbitrary constants of integration and $%
_{2}F_{1}[a,b;c,z]=\sum_{k=0}^{\infty }{\left[ (a)_{k}(b)_{k}/(c)_{k}\right]
z^{k}/k!}$ is the hypergeometric function.

\section{Numerical analysis of the reaction--convection--diffusion equations
with arbitrary power law dependencies of the diffusion and reaction functions%
}

\label{sect4}

In the present Section we present a numerical analysis of two of the exact
travelling wave solutions of the reaction--convection--diffusion equations
obtained in the previous Sections. We will restrict our analysis to the case
of the two generalized reaction--convection--diffusion equations with power
law forms of the diffusion and reaction functions, which may have a number
of important applications in chemistry, genetics and biology.

\subsection{The reaction--convection--diffusion equation with power law
diffusion and reaction functions}

We will investigate first the properties of the travelling wave solutions in
the reaction--convection--diffusion equation with power law diffusion and
reaction functions, given by
\begin{equation}
\frac{\partial u(x,t)}{\partial t}=\frac{\partial }{\partial x}\left\{ \frac{%
D_{0}}{\left[ 1-u(x,t)/u_{max}\right] ^{\alpha }}\frac{\partial u(x,t)}{%
\partial x}\right\} +B_{0}\frac{\partial u(x,t)}{\partial x}+\rho u(x,t)%
\left[ 1-\frac{u(x,t)}{u_{max}}\right] ^{\alpha },  \label{FK1}
\end{equation}%
where $u_{max}$, $\alpha $, $D_{0}$, $B_{0}$ and $\rho $ are constants, and
whose travelling wave solutions are given by Eqs.~(\ref{mod1}) and (\ref%
{mod2}), respectively.

In order to simplify Eq.~(\ref{FK1}) we introduce a set of dimensionless
variables $\left(T,X,U\right)$, defined as
\begin{equation*}
T=\frac{B_0^2}{D_0}t, X=\frac{B_0}{D_0}x,
\end{equation*}
and
\begin{equation*}
U=\frac{u}{u_{max}},
\end{equation*}
respectively. We also denote $\lambda =\rho D_0/B_0^2$. In the new variables
Eq.~(\ref{FK1}) takes the form
\begin{equation}  \label{108}
\frac{\partial U(X,T)}{\partial T}=\frac{\partial}{\partial X}\left[\frac{1}{%
(1-U)^{\alpha }}\frac{\partial U(X,T)}{\partial X}\right]+\frac{\partial
U(X,T)}{\partial X}+\lambda U(X,T)\left[1-U(x,t)\right]^{\alpha }=0.
\end{equation}

We look for solutions of Eq.~(\ref{108}) of the form $U(X,T)=U\left(
X-VT\right) =U(\xi )$, where $V$ is a dimensionless constant. Then Eq.~(\ref%
{108}) takes the form
\begin{equation}
\frac{d^{2}U}{d\xi ^{2}}+\frac{\alpha }{1-U}\left( \frac{dU}{d\xi }\right)
^{2}+(V+1)(1-U)^{\alpha }\frac{dU}{d\xi }+\lambda U(1-U)^{2\alpha }=0.
\label{109}
\end{equation}%
The successive transformations $\sigma =dU/d\xi $, $\sigma =1/v$, $%
v=(1-U)^{-\alpha }w$ reduce Eq.~(\ref{109}) to the first order first kind
Abel equation,
\begin{equation}\label{110}
\frac{dw}{dU}=(V+1)w^{2}+\lambda Uw^{3},
\end{equation}%
which, according to the Chiellini lemma, can be exactly integrated. By
introducing a new variable $\theta $ so that $w=\frac{V+1}{\lambda U}\theta $%
, then Eq. ~(\ref{110}) becomes
\begin{equation*}
kU\frac{d\theta }{dU}=\theta \left( \theta ^{2}+\theta +k\right) ,
\end{equation*}%
where $k=\lambda /(V+1)^{2}$.

Therefore Eq.~(\ref{109}) has the general solution
\begin{equation}  \label{111}
U(\theta ,k)=C_{0}^{-1}(k)e^{F(\theta ,k)},
\end{equation}
\begin{equation}
\xi (\theta ,k)-\xi _{0}\left( \theta _{0},k\right) =\frac{1}{V+1}%
\int_{\theta _{0}}^{\theta }{\frac{d\theta }{\left[ 1-C_{0}^{-1}e^{F(\theta
,k)}\right] ^{\alpha }\left( \theta ^{2}+\theta +k\right) }},  \label{112}
\end{equation}%
where $C_{0}$ and $\xi _{0}\left( \theta _{0},k\right) $ are the arbitrary
integration constants.

We consider Eq.~(\ref{109}) with the initial conditions $U(0)=U_{0}$ and $%
\left. \left( dU/d\xi \right) \right\vert _{\xi =0}=U_{0}^{\prime }$,
respectively. We assume that for $\theta =\theta _{0}$, $\xi \left( \theta
_{0},k\right) =\xi _{0}\left( \theta _{0},k\right) =0$. Then the integration
constant $C_{0}$ can be obtained as $C_{0}=e^{F\left( \theta _{0},k\right)
}/U_{0}$. Then by using the relations
\begin{equation*}
\left. \frac{dU}{d\xi }\right\vert _{\xi =0}=\left. \frac{dU}{d\theta }\frac{%
d\theta }{d\xi }\right\vert _{\theta =\theta _{0}}=\left. \frac{%
k(V+1)U\left( 1-U\right) ^{\alpha }}{\theta }\right\vert _{\theta =\theta
_{0}}=U_{0}^{\prime },
\end{equation*}
we obtain the initial value of the parameter $\theta _{0}$ as
\begin{equation*}
\theta _{0}=\frac{k(V+1)U_{0}\left( 1-U_{0}\right) ^{\alpha }}{U_{0}^{\prime
}}.
\end{equation*}

In the following we will study the properties of the solution given by Eqs.~(%
\ref{111}) and (\ref{112}) for different values of $k$.

\subsubsection{The case k=1/4}

In the case $k=1/4$, or, equivalently, $V=2\sqrt{\lambda }-1$, the general
solution of Eq.~(\ref{109}) is obtained as
\begin{equation*}
U(\theta )=C_{0}^{-1}\frac{\theta }{1+2\theta }e^{1/(1+2\theta )},
\end{equation*}
\begin{equation*}
\xi (\theta )=\frac{4}{V+1}\int_{\theta _{0}}^{\theta }\left[ 1-C_{0}^{-1}%
\frac{\theta }{1+2\theta }e^{1/(1+2\theta ) }\right] ^{-\alpha }\frac{d\theta
}{(1+2\theta )^{2}}.
\end{equation*}

The parametric dependence of $\xi $ can be obtained in an exact form for a
number of values of $\alpha $. Thus, for example,
\begin{equation*}
\xi (\theta )=\frac{4}{V+1}\left. \frac{e^{\frac{1}{2\theta +1}}(4\theta
+1)-2C_{0}}{4C_{0}(2\theta +1)}\right\vert _{\theta _{0}}^{\theta },\alpha
=-1,
\end{equation*}%
\begin{equation*}
\xi (\theta )=-\frac{4}{V+1}\left. \frac{16C_{0}^{2}(2\theta +1)-16C_{0}e^{%
\frac{1}{2\theta +1}}\left( 8\theta ^{2}+6\theta +1\right) +e^{\frac{2}{%
2\theta +1}}\left[ 4\theta (5\theta +2)+1\right] }{32C_{0}^{2}(2\theta
+1)^{2}}\right\vert _{\theta _{0}}^{\theta },\alpha =-2.
\end{equation*}

In the limit of small $\theta $, so that $2\theta <<1$ and $e\theta /C_0<<1$%
, the general solution of Eq.~(\ref{109}) for $k=1/4$ can be approximated as
\begin{equation*}
U(\xi)\approx C_0^{-1}e\theta, \xi(\theta )\approx \frac{4}{V+1}\left(\theta
-\theta _0\right),
\end{equation*}
giving
\begin{equation*}
U(\xi)\approx \left[\frac{(V+1)C_0^{-1}e}{4}\right]\xi +C_0^{-1}e\theta _0,
\end{equation*}
that is, a linear dependence of $U$ on $\xi$. In the limit of the large
values of the parameter $\theta $ we obtain $U(\theta)\approx C_0^{-1}/2$
and
\begin{equation*}
\xi (\theta)\approx \left[\frac{\left(1-C_0^{-1}/2\right)^{-\alpha }}{(V+1)}%
\right]\left.\left(-1/\theta\right)\right|_{\theta _0}^{\theta}\approx \left[%
\frac{\left(1-C_0^{-1}/2\right)^{-\alpha }}{(V+1)\theta _0}\right].
\end{equation*}
Therefore, in the limit of the large values of the parameter $\theta$, $U$
tends to a constant value.

\begin{figure}[tbp]
\centering
\includegraphics[width=8cm]{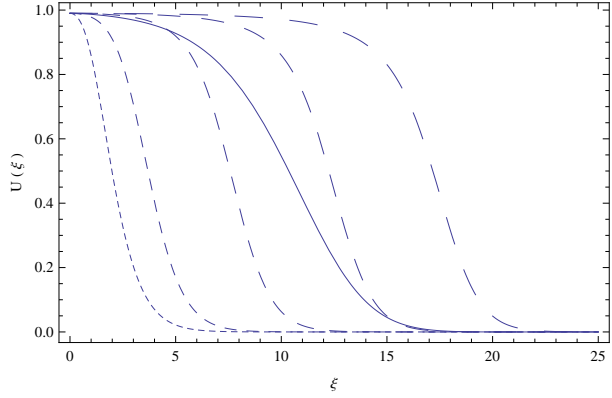}
\caption{Variation of the travelling wave solution $U$ of the  reaction--convection--diffusion equation with power law
diffusion and reaction functions as a function of $\protect\xi $ for $%
k=1/4$, and for different values of $\protect\alpha $: $\protect\alpha =1/4$
(dotted curve), $\protect\alpha =1/2$ (short dashed curve), $\protect\alpha =3/4$
(dashed curve), $\protect\alpha =0.9$ (long dashed curve) and $\protect%
\alpha =1$ (ultra long dashed curve), respectively. The initial conditions are $%
U(0)=1$ and $U^{\prime }(0)=0$, respectively. For the sake of comparison the travelling wave solution of the Fisher-Kolmogorov Eq.~(\ref{fk}) is also presented in the solid curve.}
\label{fig1}
\end{figure}

The variation of $U$ as a function of $\xi $ is represented, for different
values of $\alpha $, in Fig.~\ref{fig1}.

In order to compare our results with previously obtained travelling wave solutions in Fig.~\ref{fig1} we also present the travelling wave solution of the Fisher--Kolmogorov Eq.~(\ref{fk}). As one can see from the figure, in all considered cases the travelling waves connect the stationary states $U=1$ and $U=0$ of the underlying reaction--convection--diffusion partial differential equations.

\subsubsection{The case $k\neq 1/4$}

If the dimensionless velocity of the travelling wave has the value $V=\sqrt{%
\lambda /k}-1$, $k\neq 1/4$, the general travelling wave solution of Eq.~(%
\ref{108}) is obtained as
\begin{equation*}
U(\theta ,k)=C_{0}^{-1}\frac{\theta e^{-\frac{\tan ^{-1}\left( \frac{2\theta
+1}{\sqrt{4k-1}}\right) }{\sqrt{4k-1}}}}{\sqrt{\theta ^{2}+\theta +k}},k\neq
\frac{1}{4},
\end{equation*}
\begin{equation*}
\xi \left( \theta ,k\right) =\frac{1}{V+1}\int_{\theta _{0}}^{\theta }\frac{%
\left[ \frac{\sqrt{\theta ^{2}+\theta +k}}{\theta }-C_{0}^{-1}e^{-\frac{\tan
^{-1}\left( \frac{2\theta +1}{\sqrt{4k-1}}\right) }{\sqrt{4k-1}}}\right]
^{-\alpha }d\theta }{\theta ^{\alpha }\left( \theta ^{2}+\theta +k\right)
^{1-\alpha /2}}, k\neq \frac{1}{4}.
\end{equation*}

\begin{figure}[!]
\centering
\includegraphics[width=8cm]{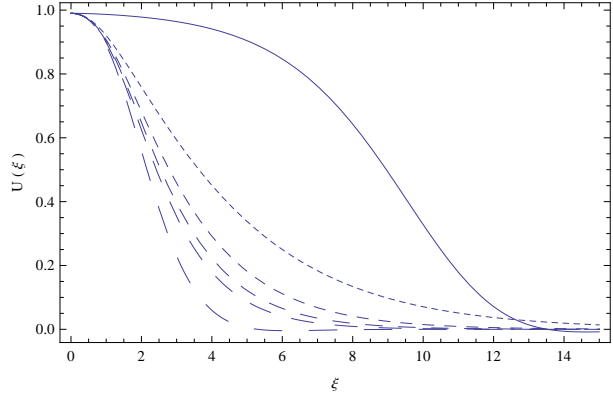}
\caption{Variation of the travelling wave solution $U$ of the  reaction--convection--diffusion equation with power law
diffusion and reaction functions as a function of $\protect\xi $ for $%
\alpha =1/4$, and for different values of $k$: $k =1/16$ (dotted curve), $k =1/8$ (short dashed curve), $k=1/6$ (dashed curve), $k=1/5$ (long dashed
curve) and $k=1/3$ (ultra long dashed curve), respectively. The initial conditions
are $U(0)=1$ and $U^{\prime }(0)=0$, respectively. For the sake of comparison the travelling wave solution of the Fisher-Kolmogorov Eq.~(\ref{fk}) is also presented in the solid curve.}
\label{fig2}
\end{figure}

The variation of $U$ as a function of $\xi$ is represented, for a fixed $%
\alpha $, and for different values of $k$, in Fig.~\ref{fig2}. In Fig.~\ref{fig2} we have also presented the solution of the Fisher--Kolmogorov equation Eq.~(\ref{fk}), corresponding to the same initial conditions.

In the limit of the large values of the parameter $\theta $, $U$ tends to
constant values given by
\begin{equation*}
\lim_{\theta \rightarrow \infty }U(\theta ,k)=C_{0}^{-1}\exp \left( -\frac{%
\pi }{2\sqrt{4k-1}}\right) ,k>1/4,
\end{equation*}%
and
\begin{equation*}
\lim_{\theta \rightarrow \infty }U(\theta ,k)=C_{0}^{-1}\exp \left( -i\frac{%
\pi }{2\sqrt{1-4k}}\right) ,k<1/4,
\end{equation*}
respectively. For small values of the parameter $\theta $ we obtain an
approximate parametric representation of the solution of the
generalized of the reaction--convection--diffusion equation with power law dependence of the diffusion and reaction functions for $k\neq 1/4$ as
\begin{eqnarray*}
C_{0}U(\theta ,k) &\approx &\frac{\theta _{0}e^{-\frac{\tan ^{-1}\left(
\frac{2\theta _{0}+1}{\sqrt{4k-1}}\right) }{\sqrt{4k-1}}}}{\sqrt{\theta
_{0}^{2}+\theta _{0}+k}}+\frac{ke^{-\frac{\tan ^{-1}\left( \frac{2\theta
_{0}+1}{\sqrt{4k-1}}\right) }{\sqrt{4k-1}}}}{\left( \theta _{0}^{2}+\theta
_{0}+k\right) ^{3/2}}\left( \theta -\theta _{0}\right) -  \notag \\
&&\frac{1}{2\left( \theta _{0}^{2}+\theta _{0}+k\right) ^{5/2}}\left[
(3\theta _{0}+2)ke^{-\frac{\tan ^{-1}\left( \frac{2\theta _{0}+1}{\sqrt{4k-1}%
}\right) }{\sqrt{4k-1}}}\right] \left( \theta -\theta _{0}\right)
^{2}+O\left( (\theta -\theta _{0})^{3}\right) ,  \notag \\
&&
\end{eqnarray*}%
and
\begin{eqnarray*}
\xi (\theta ,k) &\approx &\frac{1}{V+1}\frac{1}{\theta _{0}^{2}+\theta _{0}+k%
}\left( 1-\frac{\theta _{0}e^{-\frac{\tan ^{-1}\left( \frac{2\theta _{0}+1}{%
\sqrt{4k-1}}\right) }{\sqrt{4k-1}}}}{C_{0}\sqrt{\text{$\theta $}%
_{0}^{2}+\theta _{0}+k}}\right) ^{-\alpha }\times  \notag \\
\ &&\left\{ \left( \theta -\theta _{0}\right) -\frac{\left[ \alpha
k+(2\theta _{0}+1)\left( \theta _{0}-C_{0}\sqrt{\theta _{0}^{2}+\theta _{0}+k%
}e^{\frac{\tan ^{-1}\left( \frac{2\theta _{0}+1}{\sqrt{4k-1}}\right) }{\sqrt{%
4k-1}}}\right) \right] }{2\left( \theta _{0}^{2}+\theta _{0}+k\right) \left[
\theta _{0}-C_{0}\sqrt{\theta _{0}^{2}+\theta _{0}+k}e^{\frac{\tan
^{-1}\left( \frac{2\theta _{0}+1}{\sqrt{4k-1}}\right) }{\sqrt{4k-1}}}\right]
}(\theta -\theta _{0})^{2}\right\} +  \notag \\
&&O\left( (\theta -\theta _{0})^{3}\right) ,
\end{eqnarray*}%
respectively.

\subsection{The reaction--convection--diffusion equation with power law
reaction function}

We consider now a mathematical model in which the diffusion and reaction
functions satisfy the conditions given by Eq.~(\ref{d4}). Moreover, we
assume $D(u)=D_0=\mathrm{constant}$, which fixes the reaction function as
\begin{equation*}
Q(u)=\frac{2(m+1)}{(m+3)^2}\frac{B_0^2}{D_0}\lambda u-Au^m,m\neq-1,m\neq -3,
\end{equation*}
where $A$ and $\lambda $ are constants. The corresponding
reaction--convection--diffusion equation is given by
\begin{eqnarray}  \label{95}
\frac{\partial u(x,t)}{\partial t}&=&D_0\frac{\partial ^2u(x,t)}{\partial x
^2}+B_0\frac{\partial u(x,t)}{\partial x} +\frac{2(m+1)}{(m+3)^2}\frac{B_0^2%
}{D_0}\lambda u(x,t)-Au^m(x,t),  \notag \\
&&m\neq-1,m\neq -3.
\end{eqnarray}

We call Eq.~(\ref{95}) \textit{the reaction--convection--diffusion equation
with power law reaction function}. By introducing the set of dimensionless
variables $\left( T,X,U\right) $, defined as
\begin{equation*}
T= \frac{B_{0}^{2}}{D_{0}} t,X= \frac{B_{0}}{D_{0}} x,
\end{equation*}
and
\begin{equation*}
U=\left( \frac{B_{0}^{2}}{AD_{0}}\right)^{-1/(m-1)}u,
\end{equation*}
respectively, Eq.~(\ref{95}) takes the form
\begin{equation}
\frac{\partial U(X,T)}{\partial T}=\frac{\partial ^{2}U(X,T)}{\partial X^{2}}%
+\frac{\partial U(X,T)}{\partial X}+\frac{2(m+1)}{(m+3)^{2}}\lambda
U(X,T)-U^{m}(X,T).  \label{123}
\end{equation}

We look for travelling wave solutions of Eq.~(\ref{123}), so that $%
U(X,T)=U(\xi )=U(X-VT)$. Therefore Eq.~(\ref{123}) becomes
\begin{equation}
\frac{d^{2}U(\xi )}{d\xi ^{2}}+(V+1)\frac{dU(\xi )}{d\xi }+\frac{2(m+1)}{%
(m+3)^{2}}\lambda U(\xi )-U^{m}(\xi )=0.  \label{124}
\end{equation}%
At this moment we rescale the parameter $\xi $, so that $\xi \rightarrow
\frac{\xi }{V+1}$. Moreover, we fix $\lambda $ to the value $\lambda
=(V+1)^{2}$. Therefore Eq.~(\ref{124}) becomes
\begin{equation}
\frac{d^{2}U(\xi )}{d\xi ^{2}}+\frac{dU(\xi )}{d\xi }+\frac{2(m+1)}{(m+3)^{2}%
}U(\xi )-\frac{1}{(V+1)^{2}}U^{m}(\xi )=0.  \label{125}
\end{equation}%
Using the substitution $v=d\xi /dU$, thus we write Eq.~(\ref{125}) as an
Abel equation in the form
\begin{equation*}
\frac{dv}{dU}=v^{2}+\left[ \frac{2(m+1)}{(m+3)^{2}}U-\frac{1}{(V+1)^{2}}U^{m}%
\right] v^{3}.
\end{equation*}%
Finally, by taking $v=-(1/w)(dw/dU)$, it follows that $U$ can be obtained as
a solution of the second order differential equation
\begin{equation}
w^{2}\frac{d^{2}U}{dw^{2}}+\left[ \frac{2(m+1)}{(m+3)^{2}}U-\frac{1}{%
(V+1)^{2}}U^{m}\right] =0.  \label{127}
\end{equation}

The general solution of Eq.~(\ref{127}) is given, in a parametric form, by
\begin{equation*}
U(\theta )=b\theta \left[ \theta \,_{2}F_{1}\left( \frac{1}{2},\frac{1}{m+1}%
;1+\frac{1}{m+1};-\theta ^{m+1}\right) +l_{2}\right] {}^{\frac{2}{m-1}},
\end{equation*}%
\begin{equation*}
w(\theta )=l_{1}\left[ \theta \,_{2}F_{1}\left( \frac{1}{2},\frac{1}{m+1};1+%
\frac{1}{m+1};-\theta ^{m+1}\right) +l_{2}\right] {}^{\frac{m+3}{m-1}},
\end{equation*}%
where
\begin{equation*}
b=\left[ \frac{(m-1)^{2}(m+1)(V+1)^{2}}{2(m+3)^{2}}\right] ^{1/(m-1)},
\end{equation*}%
and $l_{1}$ and $l_{2}$ are the arbitrary constants of integration. The
parametric dependence of $\xi $ is obtained as
\begin{equation*}
\xi (\theta )-\xi _{0}=-\frac{m+3}{m-1}\ln \left[ \theta \,_{2}F_{1}\left(
\frac{1}{2},\frac{1}{m+1};1+\frac{1}{m+1};-\theta ^{m+1}\right) +l_{2}\right]
,
\end{equation*}%
where $\xi _{0}$ is an arbitrary constant of integration. The integration
constant $l_{2}$ and the initial value of the parameter $\theta _0=\theta (0)$
can be obtained from the initial conditions $\left. U(\xi )\right\vert _{\xi =0}=U_{0}$ and $%
\left. \left[ dU(\xi )/d\xi \right] \right\vert _{\xi =0}=U_{0}^{\prime }$
by simultaneously solving the equations
\begin{equation*}
U_{0}=b\theta _{0}\left[ \theta _{0}\,_{2}F_{1}\left( \frac{1}{2},\frac{1}{%
m+1};1+\frac{1}{m+1};-\theta _{0}^{m+1}\right) +l_{2}\right] {}^{\frac{2}{m-1%
}},
\end{equation*}%
\begin{eqnarray*}
U_{0}^{\prime } &=&-\frac{b}{m+3}\left[ \theta _{0}\,\,_{2}F_{1}\left( \frac{%
1}{2},\frac{1}{m+1};1+\frac{1}{m+1};-\theta _{0}^{m+1}\right) +l_{2}\right]
^{\frac{2}{m-1}}\times  \\
&&\left[ 2\theta _{0}+\theta _{0}(m-1)\sqrt{\theta _{0}^{m+1}+1}%
\,_{2}F_{1}\left( \frac{1}{2},\frac{1}{m+1};1+\frac{1}{m+1};-\theta
_{0}^{m+1}\right) +l_{2}(m-1)\sqrt{\theta _{0}^{m+1}+1}\right] . \\
&&
\end{eqnarray*}

The integration constant $\xi _0$ can be obtained from the equation $%
U_0=b\theta _0\exp \left[2\xi _0/(m+3)\right]$. The variation of $U$ as a
function of $\xi $ is represented, for different values of $m$, and for $V=2$, in Fig.~\ref%
{fig3}. For the sake of comparison the travelling wave solution of the Fisher--Kolmogorov equation Eq.~(\ref{fk}), corresponding to the same initial conditions and the same value of $V$ is also presented.

\begin{figure}[!]
\centering
\includegraphics[width=8cm]{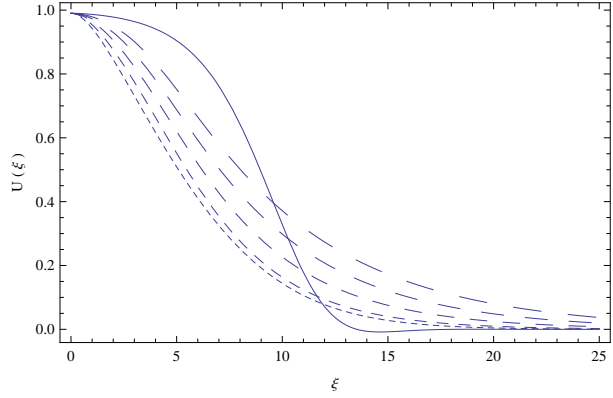}
\caption{Variation of the travelling wave solution $U$ of the  reaction--convection--diffusion equation with constant diffusion coefficient and power law
reaction function as a function of $\protect\xi $  for different values of $m$: $m =2$ (dotted curve), $m
=4 $ (short dashed curve), $m=6$ (dashed curve), $m=8$ (long dashed curve)
and $m=10$ (ultra long dashed curve), respectively. The initial conditions are $%
U(0)=1$ and $U^{\prime }(0)=0$, respectively. For the sake of comparison the travelling wave solution of the Fisher-Kolmogorov Eq.~(\ref{fk}) is also presented in the solid curve.}
\label{fig3}
\end{figure}

\section{Discussions and final remarks}

\label{sect5}

In the present paper we have investigated the travelling wave solutions of a
general class of reaction--convection--diffusion systems, with the
diffusion, convection and reaction functions given as functions of the
concentration $u$ of the diffusing--convecting--reacting component. In this
case the second order differential equation, describing the travelling wave
solution for the initial second order non-linear partial differential
equation, can be reduced to a first kind first order Abel differential
equation. The integrability conditions of the Abel equation allow to obtain
several classes of exact travelling wave solutions of the general
reaction--convection--diffusion system, with the reaction, convection and
diffusion functions $Q(u)$, $B(u)$ and $D(u)$, respectively, satisfying some
functional integrability conditions. We have considered the solutions that
can be found by using the Chiellini integrability condition of the Abel
equation, and those that follow from the Lemke transformation of the Abel
equation, respectively. The use of these mathematical results allows
obtaining a large number of travelling wave solutions of Eq.~(\ref{1}). In
particular, from the many exactly integrable reaction--convection--diffusion
equations, we did concentrate on two equations, with power law type
diffusion and reaction functions. More exactly, with the choices
\begin{equation*}
D(u)=\frac{D_{0}}{\left( 1-u/u_{max}\right) ^{\alpha }},Q(u)=\rho u\left( 1-%
\frac{u}{u_{max}}\right) ^{\alpha },B(u)=\mathrm{constant},
\end{equation*}%
and
\begin{equation*}
D(u)=D_{0}=\mathrm{constant},Q(u)=\frac{2m+1}{(m+3)^{2}}\frac{\left(
V_{f}+B_{0}\right) ^{2}}{D_{0}}c\left[ 1-A\left( \frac{u}{u_{max}}\right)
^{m-1}\right] ,B(u)=\mathrm{constant},
\end{equation*}%
for the diffusion, reaction and convection functions, respectively, the
travelling wave solutions of the corresponding one dimensional
reaction--convection--diffusion equations can be obtained in an exact parametric form.

Both equations represent generalizations of the standard diffusion and
Fisher--Kolmogorov equations, to which they reduce in some limiting cases.
For the reaction--convection--diffusion equation with power law type diffusion
and reaction functions, for $\alpha =0$ we reobtain the standard diffusion
equation, while for $m=2$ the reaction--convection--diffusion equation with
power law reaction function reduces to the Fisher--Kolmogorov equation. Both
these equations have many important biological and physical applications.

An important result related to the study of the Fisher--Kolmogorov equation (\ref{fk}) is the estimation of the minimum travelling wave speed as $V_{min}=2\sqrt{\rho D}$, where $D$ and $\rho $ are the diffusion and the linear growth rate, respectively \cite{13}. For sufficiently sharp initial fronts in $u(x,0)$, the minimum speed is selected. More exactly, for sharp initial fronts any reasonable definition of the instantaneous velocity $v(t)$ satisfies $V(t)=V_{min}+O\left(t^{-1}\right)$ \cite{front}. It is interesting to  note that the Chiellini integrability condition automatically fixes the relation between the wave velocity $V_f$ and the physical model parameters as
\begin{equation}\label{vel}
V_f+B_0=\pm\sqrt{\frac{\rho D_0}{k}}.
\end{equation}
For the particular case $k=1/4$ and $B_0=0$ (corresponding to the absence of convection), we reobtain the minimum speed of the travelling waves of the Fisher--Kolmogorov equation, $V_f=V_{min}$. Therefore, by analogy with the case of the Fisher--Kolmogorov equation,  we may conjecture that for sufficiently sharp initial conditions, for all models considered in the present paper the wave front speed $V$ satisfies the relation $V=V_f+O\left(t^{-1}\right)$. A rigorous investigation of the speed propagation of the travelling waves in the general reaction--convection--diffusion equation Eq.~(\ref{1}) would require the phase plane analysis of the corresponding one-dimensional system,
\begin{equation*}
\frac{du}{d\xi}=\Psi,
\end{equation*}
\begin{equation*}
\frac{d\Psi }{d\xi }=-\alpha (u)\Psi ^2-\beta (u)\Psi -\gamma (u).
\end{equation*}

The fixed points $\left(u_c,\Psi_c\right)$ of the above system of differential equations are the points where $\Psi =0$ and $\gamma \left(u_c\right)=0$. They correspond to the steady states of the general reaction--convection--diffusion equation. The local behaviour of the trajectories of the first order equivalent system  can be obtained by analyzing the linear approximation of the system around each fixed point. The stability of the fixed points gives the speed of the travelling wave, as well as its relation to the minimum speed \cite{13}. Such an analysis can be performed on a case by case basis, once the explicit form of the function $\gamma (u)=Q(u)/D(u)$ is known.

Most of the mathematical studies of the reaction--convection--diffusion
equations have been done by using either qualitative methods, or asymptotic
evaluations. The usual way to study nonlinear reaction--diffusion equations
similar to Eq.~(\ref{3}) consists of the analysis of the behavior of the
system on a phase plane $(du/d\xi, u)$, useful for qualitative analysis,
however insufficient for finding any exact solution, or testing a numerical
one. Finding exact analytical solutions of the complex
reaction--convection--diffusion models can considerably simplify the process
of comparison of the theoretical predictions with the experimental data,
without the need of using complicated numerical procedures.

\section*{Acknowledgments}

We would like to thank to the anonymous referee for comments and suggestions
that helped us to significantly improve our manuscript.


\begin{thebibliography}{99}
\bibitem{9cm} B. H. Gilding and R. Kersner, The characterization of
reaction-convection-diffusion processes by travelling waves, Journal of
differential equations \textbf{124}, 27--79 (1996).

\bibitem{b1} Z. Mei, Numerical bifurcation analysis for reaction-diffusion
equations, Springer-Verlag Berlin Heidelberg New York, 2000

\bibitem{b2} H. Wilhelmsson and E. Lazzaro, Reaction-diffusion problems in
the physics of hot plasmas, Institute of Physics Publishing, Dirac House,
Temple Back, Bristol, United Kingdom, 2001.

\bibitem{b3} B. H. Gilding and R. Kersner, Travelling waves in nonlinear
diffusion convection reaction, Birkh\"{a}user, 2004

\bibitem{b4} Y. Du, H. Ishii, and W.-Y. Lin (Editors), Recent progress on
reaction-diffusion systems and viscosity solutions, World Scientific
Publishing Co., Singapore, 2009.

\bibitem{b5} A. W. Liehr, Dissipative solitons in reaction diffusion
systems, mechanisms, dynamics, interaction, Springer-Verlag Berlin
Heidelberg, 2013.

\bibitem{Fish} R. A. Fisher, The genetical theory of natural selection,
Oxford, Oxford University Press, 2000.

\bibitem{Fish1} R. A. Fisher, The wave of advance of advantageous genes,
Ann. Eugenics \textbf{7}, 353--369, (1937).

\bibitem{Kolm} A. Kolmogorov, I. Petrovskii, and N. Piscounov, A study of
the diffusion equation with increase in the amount of substance, and its
application to a biological problem, Bull. Moscow Univ., Math. Mech. \textbf{%
1}, 1--25, (1937).

\bibitem{13} J. D. Murray, Mathematical Biology, 3rd ed, New York, NY:
Springer-Verlag, 2002.

\bibitem{N_W} A. C. Newell and J. A. Whitehead, Finite bandwidth, finite
amplitude convection, Journal of Fluid Mechanics \textbf{38}, 279--303
(1969).

\bibitem{S} L. A. Segel, Distant side-walls cause slow amplitude modulation
of cellular convection, Journal of Fluid Mechanics \textbf{38}, 203--224,
(1969).

\bibitem{Zel} Y. B. Zeldovich and D. A. Frank--Kamenetsky, A theory of
thermal propagation of flame, in Dynamics of Curved Fronts (edited by P.
Pelce), Academic Press, Boston, 1988, pp. 131-140. Translation from: Acta
Physicochimica U.R.S.S. \textbf{9}, 124--131 (1938).

\bibitem{Zel1} Y. B. Zeldovich, Theory of flame propagation, National
Advisory Committee for Aeronautics Technical Memorandum 1282 (1951), 39 pp.
Translation of: Zh. Fiz. Khim. \textbf{22}, 27--49 (1948).

\bibitem{Fitz} R. FitzHugh, Impulses and physiological states in theoretical
models of nerve membrane, Biophysical Journal \textbf{1}, 445--466 (1961).

\bibitem{Nag} J. Nagumo, S. Arimoto, and S. Yoshizawa, An active pulse
transmission line simulating nerve axon, Proc. Inst. Radio Engin. Electr.
\textbf{50}, 2061--2070, (1962).

\bibitem{9c1} J. D. Murray, On travelling wave solutions in a model for the
Belousov-Zhabotinskii reaction, Journal of Theoretical Biology \textbf{56},
329--353 (1976).

\bibitem{9c01} M. Bellini, R. Deza, and N. Giovanbattista, Exact travelling
annular waves in generalized reaction-diffusion equations, Physics Letters
\textbf{A 232}, 200--206 (1997).

\bibitem{9c2} A. M. Samsonov, On exact quasistationary solutions to a
nonlinear reaction-diffusion equation, Physics Letters \textbf{A 245},
527--536 (1998).

\bibitem{Golub} V. V. Golubev, Vorlesungen \"{u}ber Differentialgleichungen
im Komplexen, Deutsch. Verlag Wissenschaft., Berlin, 1958.

\bibitem{9c3} T. Brugarino, Exact solutions of nonlinear differential
equations using the Abelian equation of the first type, Il Nuovo Cimento
\textbf{119 B}, 975--982 (2004).

\bibitem{9c4} S. C. Mancas, G. Spradlin, and H. Khanal, Weierstrass
traveling wave solutions for dissipative Benjamin, Bona, and Mahony (BBM)
equation, Journal of Mathematical Physics \textbf{54}, 081502 (2013).

\bibitem{9c} K. R. Swanson and E. C. Alvord Jr, The concept of gliomas as a
travelling wave - application of a mathematical model to high - and
low-grade gliomas, Can. J. Neurol. Sci. \textbf{29}, 395 (2002).

\bibitem{9d} D. C. Markham, M. J.Simpson, P. K. Maini, E. A.Gaffney, and R.
E. Baker, Comparing methods for modelling spreading cell fronts, Journal of
Theoretical Biology \textbf{353}, 95-103 (2014).

\bibitem{10} A. Matzavinos and M. A.J. Chaplain, Traveling-wave analysis of
a model of the immune response to cancer, C. R. Biologies \textbf{327}, 995
- 1008 (2004).

\bibitem{11} P. Gerlee and S. Nelander, Travelling wave analysis of a
mathematical model of glioblastoma growth, arXiv:1305.5036 (2013).

\bibitem{12} B. Perthame, M. Tang, and N. Vauchelet, Traveling wave solution
of the Hele-Shaw model of tumor growth with nutrient, Mathematical Models
and Methods in Applied Sciences \textbf{24}, 2601--2626 (2014).

\bibitem{H14} T. Harko and M. K. Mak, Travelling wave solutions of the
reaction-diffusion mathematical model of glioblastoma growth: An Abel
equation based approach, Mathematical Biosciences and Engineering \textbf{12}%
, 41--69 (2015).

\bibitem{kamke} E. Kamke, Differentialgleichungen: L\"{o}sungsmethoden und L%
\"{o}sungen, Chelsea, New York, 1959.

\bibitem{18} M. K. Mak, H. W. Chan, and T. Harko, Solutions generating
technique for Abel type non-linear ordinary differential equations, Computer
and Mathematics with Applications \textbf{41}, 1395-1401 (2001).

\bibitem{17} M. K. Mak and T. Harko, New method for generating general
solution of Abel differential equations, Computer and Mathematics with
Applications \textbf{43}, 91-94 (2002).

\bibitem{16} T. Harko and M. K. Mak, Relativistic dissipative cosmological
models and Abel differential equation, Computer and Mathematics with
Applications \textbf{46}, 849-853 (2003).

\bibitem{19} T. Harko, F. S. N. Lobo, and M. K. Mak, 2014, Exact analytical
solutions of the Susceptible-Infected- Recovered (SIR) epidemic model and of
the SIR model with equal deaths and births, Applied Mathematics and
Computation \textbf{236}, 184-194 (2014).

\bibitem{H13} T. Harko, F. S. N. Lobo, and M. K. Mak, A Chiellini type
integrability condition for the generalized first kind Abel differential
equation, Universal Journal of Applied Mathematics \textbf{1}, 101-104
(2013).

\bibitem{Rosu1} S. C. Mancas and H. C. Rosu, Integrable dissipative
nonlinear second order differential equations via factorizations and Abel
equations, Physics Letters, \textbf{A 377}, 1234--1238 (2013).

\bibitem{Rosu2} S. C. Mancas and H. C. Rosu, Integrable equations with
Ermakov-Pinney nonlinearities and Chiellini damping, Applied Mathematics and
Computation \textbf{259}, 1--11 (2015).

\bibitem{HLM} T. Harko, F. S. N. Lobo, and M. K. Mak, A class of exact
solutions of the Li\'{e}nard type ordinary non-linear differential equation,
Journal of Engineering Mathematics \textbf{89}, 193--205 (2014).

\bibitem{Rosu3} H. C. Rosu, S. C. Mancas, and P. Chen, Barotropic FRW
oscillators with Chiellini damping, Phys. Lett. \textbf{A 379}, 882--887
(2015).

\bibitem{Rosu4} S. C. Mancas and H. C. Rosu, Integrable Abel equations and
Vein's Abel equation, arXiv:1505.03548 (2015).

\bibitem{Chiel} A. Chiellini, Sull'integrazione dell'equazione differenziale
$y^{\prime }+Py^{2}+Qy^{3}=0$, Bollettino dell Unione Matematica Italiana,
10, 301 - 307 (1931).

\bibitem{Lemke} H. Lemke, \"{U}ber eine von R. Liouville untersuchte
Differentialgleichung erster Ordnung, Sitzungs. Berl. Math. Ges. 18, 26-31,
(1920).

\bibitem{Pol} A. D. Polyanin and V. F. Zaitsev, Handbook of Exact Solutions
for Ordinary Differential Equations, Chapman \& Hall/CRC, Boca Raton London
New York Washington, D.C., 2003.

\bibitem{front} U. Ebert and W. van Saarloos, Front propagation into unstable states: universal algebraic
convergence towards uniformly translating pulled fronts, Physica {\bf D 146}, 1--99 (2000).

\end{thebibliography}
\end{document}